\documentclass[12pt]{iopart}

\usepackage{iopams}
\usepackage[T1]{fontenc}
\usepackage[utf8]{inputenc}
\usepackage[english]{babel}
\usepackage[babel]{csquotes}

\expandafter\let\csname equation*\endcsname\relax
\expandafter\let\csname endequation*\endcsname\relax
\usepackage{amsmath}
\usepackage{amsfonts, amssymb,amsxtra,amscd,amsthm}
\usepackage{amsfonts}
\usepackage{mathrsfs,ae,dsfont,algpseudocode,algorithm}
\usepackage{graphicx}
\usepackage[normalem]{ulem}
\usepackage{nicefrac,xcolor, cmll, verbatim}
\usepackage{subfigure}

\theoremstyle{definition}

\begin{document}

\title[SMC samplers for semi--linear IPs]{Sequential Monte Carlo Samplers for Semi--Linear Inverse Problems
and Application to Magnetoencephalography}

\author{Sara Sommariva and Alberto Sorrentino}
\address{Dipartimento di Matematica, Genova and CNR--SPIN, Genova}

\begin{abstract}
We discuss the use of a recent class of sequential Monte Carlo methods for solving inverse
problems characterized by a semi--linear structure, i.e. where the data depend linearly on a subset
of variables and non--linearly on the remaining ones. In this type of problems, under proper Gaussian assumptions
one can marginalize the linear variables. This means that the Monte Carlo procedure needs only to be applied
to the non--linear variables, while the linear ones can be treated analytically; as a result, the Monte Carlo variance and/or the computational cost decrease.
We use this approach to solve the inverse problem of magnetoencephalography,
with a multi--dipole model for the sources. Here, data depend non--linearly on the number of sources and
their locations, and depend linearly on their current vectors. The semi--analytic approach
enables us to estimate the number of dipoles and their location from a whole time--series, rather than
a single time point, while keeping a low computational cost.
\end{abstract}

\section{Introduction}

In the Bayesian formulation of inverse problems \cite{tava82,soka04}, the unknown and the data are
modeled as Random Variables (RVs), and the available information is coded in their probability
distributions. When the inverse problem is linear and both the prior density and the likelihood
function are Gaussian, the posterior density is also Gaussian, and it is possible
to compute analytically standard estimators like the mean and the covariance matrix. On the other
hand, characterizing the posterior distribution for non--linear/non--Gaussian problems is typically
more difficult, because analytic formulae are seldom available, and one has to resort to numerical
approximations, such as Monte Carlo sampling.

In many applied problems, the forward equation establishes a linear dependence of
the data on a subset of the unknowns, and a non--linear dependence on the remaining ones.
When the likelihood is Gaussian, this entails that the conditional distribution for the linear
variables is also normal as long as a Gaussian prior is selected for these variables.
In this paper, we discuss how to exploit such linear sub--structure when the posterior distribution
is approximated by means of a specific class of Sequential Monte Carlo (SMC) methods,
and we apply the method to the inverse source problem in Magnetoencephalography.

Indeed, the term SMC methods is widely used in the literature to refer to a varied class of Monte Carlo
algorithms, that include for instance the particle filters \cite{dogoan00,dojo11} for dynamic models.
Exploitation of a linear substructure in a particle filter leads to the well--known
Rao-Blackwellized particle filter \cite{savela07,vi07}.
Importantly, in that case the structure of the equations is such that the distributions
over the non-linear variables sampled in the Rao-Blackwellized particle filter are the marginals
of those used in the full particle filter.

In the present paper, with the term SMC \textit{samplers} we refer to a specific class of Monte Carlo algorithms, the one described in \cite{dedoja06}.
These algorithms sample a target distribution, e.g. the posterior of a Bayesian inverse problem, by constructing a sequence of
distributions that reaches it smoothly. The underlying idea is the same as that of simulated annealing
\cite{kigeve83}, however, while simulated annealing is a stochastic optimization tool, the SMC samplers
approximate the whole posterior distribution.
In a couple of recent papers \cite{waza11,soluar14} it has been proposed to use Sequential Monte
Carlo samplers for solving static inverse problems whose posterior distribution
is particularly complex and possibly multimodal. In this paper, we discuss the use of SMC samplers for semi--linear models
by means of a \textit{semi--analytic} approach: an SMC sampler is used to approximate only the marginal posterior
of the non--linear variables, then the conditional posterior of the linear ones is computed analytically.
Similar combinations of analytic calculation and SMC sampler can be found, e.g., in \cite{naasjo12}
for the case of change point problems, where the model is not semi--linear but still allows exact computation of part
of the distribution, and in \cite{gimide13}, where the structure of a conditional Hidden Markov Model
is exploited analytically with a Kalman smoother.

The first manifest advantage of the semi--analytic approach is that lesser variables are sampled, which
expectedly leads to a reduced Monte Carlo variance of the estimators.
Perhaps more importantly, this approach becomes particularly interesting when the number
of linear variables is relatively high, in that it allows to solve much larger inverse problems
than those that can be solved with a \textit{full} SMC, where sampling the full posterior becomes computationally unfeasible.
We demonstrate
this by applying the method to the inverse problem in magnetoencephalography (MEG). Here the problem is
the one to estimate the locations and the current vectors of a small set of sources, whose number is
also unknown, given a sampled magnetic field. The data depend non--linearly on the number of sources
and on their location, whereas depend linearly on the current vectors. In \cite{soluar14} a full SMC sampler
was used to analyze a single topography, i.e. a static spatial distribution of the magnetic field;
in the present paper we discuss how to improve this method by exploiting the linear dependence over the current vectors.
Moreover MEG data are sampled in time and, since the sampling frequency is usually high, source locations
can be assumed to be fixed during a time window up to hundred milliseconds. If one assumes that the number of
sources and the source locations are stationary, one can actually apply the semi--analytic approach using a whole
time--series to estimate these parameters, thus benefiting from having more signal. As we will show,
an additional assumption about the independence of different time points leads to the pleasant consequence
that the computational cost of the proposed method is dominated by calculations that do
not depend on the length of the time window.
Of course, the same Bayesian model could be approximated, in principle, also with a full SMC; however, the dimension
of the state space to be explored with Monte Carlo would increase very quickly, leading to an unbearable
computational cost. Alternatively, one could use parametric models for the source time courses, such as those
described in \cite{juetal05,juetal06}, at the price of a lower generality of the model.

Another effort towards the estimation of static dipoles from MEG time--series has been done in \cite{soetal13};
however, in that case the problem has been described as a fully dynamic problem, in which also the number of dipoles
is allowed to vary in time, and a particle filter was used to approximate dynamically the sequence of posterior distributions.
In our model the number of sources is fixed, and the only dynamic variables are the dipole moments,
but the computational cost of our new method is definitely lower. On the other hand, exploitation of the linear
substructure in the context of multi--dipole models has been performed in \cite{caetal08}, where standard
Rao--Blackellization was applied to dynamic estimation of dipoles with particle filtering.\\

The paper is organized as follows. In Sections \ref{Sec:BIPSMC} and \ref{par:SMC_clm} we recall the basic ideas behind
the use of SMC samplers for Bayesian inverse problems, and specifically for semi--linear inverse problems, respectively.
In Section \ref{par:MEG} we discuss the application of the semi--analytic method for
approximating the posterior distribution for
an unknown number of sources in magnetoencephalography, while in Section \ref{par:Simulation} and \ref{par:real_data} we validate it
with both synthetic and real MEG data. Finally our conclusions are presented in Section \ref{par:discussion}.

\section{Adaptive Sequential Monte Carlo samplers for Bayesian Inverse Problems}\label{Sec:BIPSMC}

Consider a general Bayesian inverse problem: let $Y$, $X$ and $E$ be the RVs representing the data, the unknown and the
noise, respectively; assume they are related by
\begin{equation}
Y = F(X,E)
\end{equation}
where $F$ embodies the forward model.\\

Given a realization of the data $Y = y$, the goal is to approximate the posterior density $\pi(x|y)$, which is related to the prior $\pi(x)$ and to the likelihood function $\pi(y|x)$ by the Bayes' formula
\begin{equation}\label{eq:post_pdf}
\pi(x|y) = \frac{\pi(x) \pi(y|x)}{\pi(y)}
\end{equation}
where $\pi(y)$ plays the role of the normalizing constant.

Whenever the forward model is non--linear and/or the prior and/or the likelihood are not Gaussian,
the resulting posterior distribution tends to be a rather complex function.
In these cases Monte Carlo methods are typically used to characterize such distributions, by providing
a sample set that can be used to compute estimators, such as conditional expectations, variance, and so on \cite{soka04}.

In \cite{dedoja06} Del Moral et al. presented a particular class of Monte Carlo methods,
named Sequential Monte Carlo (SMC) samplers, that result to be particularly good at approximating
complex distributions thanks to their sequential nature.\\
SMC samplers are based on two main ideas:

\begin{itemize}
\item[(i)] rather than sampling directly the complex distribution of interest,
a one--parameter family of densities is built, that transits
\emph{smoothly} from a simpler distribution to the \emph{target} one.

In the case of Bayesian inverse problems, the target distribution is the posterior (\ref{eq:post_pdf}), thus
a natural choice for such family is
\begin{equation}\label{eq:pdfs_wholeSMC}
\pi_n(x|y) = \frac{\pi(x)\ \pi(y|x)^{\alpha_n}}{\pi_n(y)}\ \ \ \ \ \ \ n = 1, \dots, N
\end{equation}
where the likelihood is raised to a tempering exponent that increases with the iterations so that $0 = \alpha_1 < \dots, < \alpha_N = 1$.
As we will describe, the values of these exponents need not to be fixed a priori but can be chosen on-line in a adaptive manner.
As observed in \cite{soluar14}, such sequence of distributions has a nice interpretation in the case of a Gaussian
likelihood: because a Gaussian density raised to an exponent is still a Gaussian density, up to a normalizing factor,
each distribution of this sequence can be interpreted as a posterior distribution for a different (decreasing) value of
the estimated noise variance.

\item[(ii)] since
direct importance sampling from this family of distributions is still not applicable, one can actually perform
importance sampling in an increasing state space by sampling from a family of artificial joint distributions whose
marginals coincide with the original ones. If $\mathcal{X}$ is the state--space for the unknown variable $x$, one can perform importance sampling on $\mathcal{X}^n =
\mathcal{X} \times \cdots \times \mathcal{X}$, defining the target joint distributions
\begin{equation}
\tilde{\pi}_n (x_{1:n}|y) = \pi_n(x_n|y) \prod_{k=2}^n L_{k-1}(x_{k}, x_{k-1})
\end{equation}
and the joint importance densities
\begin{equation}
\tilde{\eta}_n (x_{1:n}) = \eta_1(x_1) \prod_{k=2}^n K_k(x_{k-1}, x_{k})
\end{equation}
where $x_{1:n} := (x_1, \dots, x_n)$, while $\{ K_k(x_{k-1}, x_k) \}_{k=2}^N$ and $\{ L_{k-1}(x_{k}, x_{k-1}) \}_{k=2}^N$ are
two families of Markov kernels named forward and backward kernels respectively. The resulting importance weights are
naturally defined as

\begin{equation}
w_n (x_{1:n})= \frac{\tilde{\pi}_n(x_{1:n}|y)}{\tilde{\eta}_n(x_{1:n})}
\end{equation}
\end{itemize}

\noindent
Implementation of these two ideas requires proper choices of the transition kernels involved: in \cite{dedoja06}
the authors suggest that one can choose $K_k(x_{k-1}, x_k)$ to be a Markov Kernel of invariant distribution
$\pi_k(x_k|y)$ and then set \begin{equation}\label{eq:NM_scelti}
L_{k-1}(x_{k}, x_{k-1}) = \frac{\pi_{k}(x_{k-1}|y)K_k(x_{k-1},x_k)}{\pi_k(x_k|y)},
\end{equation}
that approximate the \textit{optimal backward Kernels}, i.e. the kernels that minimize the variance of the
importance weights \cite{dedoja06}. Under these assumptions the algorithm proceeds as follows. An initial set
of $I$ weighted particles $\{X_1^i, W_1^i \}_{i=1}^I$ is drawn from the first importance density $\eta_1(x_1)$
usually set equal to the first target distribution $\pi_1(x_1|y) = \pi(x_1)$. Then each particle is let evolve
according to the kernels $K_k(x_{k-1}, x_k)$ and the associated unnormalized weight is computed recursively by the formula

\begin{equation}\label{eq:recursion_weight}
w_n^i = W_{n-1}^i \pi(y|X_{n-1}^i)^{\alpha_n - \alpha_{n-1}};
\end{equation}
then  $W_n^i = \frac{w_n^i}{\sum_{j=1}^I w_n^j}$.\\
Eq. (\ref{eq:recursion_weight}) is a consequence of the choice of backward and forward kernels and of the particular
choice for the sequence of distributions (\ref{eq:pdfs_wholeSMC}). Moreover it shows that at each iteration the weight associated to the $i-$th particle depends only on the value
of the particle at the previous step, and therefore can be calculated before the new particle is drawn. This fact allows
to adaptively choose the sequence of tempering exponents for the likelihood in various ways \cite{dedoja12,soluar14, zhjoas13arxiv}.
Here, at each iteration $n$ we compute the normalized weight $W_{n+1}^i$ and the so--called Effective Sample Size (ESS),
that is an empirical measure of how much $\pi_{n+1}$ differs from $\pi_n$:
\begin{equation}
ESS(n+1) = \left( \sum_{i=1}^I (W_{n+1}^i)^2 \right)^{-1}.
\end{equation}
Then the increase $\alpha_{n+1}-\alpha_n$ is chosen by a bisection procedure that ends when the ratio $ESS(n+1) / ESS(n)$
falls into a fixed interval. Alternative procedures for this adaptive choice are described e.g. in \cite{zhjoas13arxiv}.

It is well--known that in sequential importance sampling the variance of the (unnormalized) importance weights inevitably increases as the iterations proceed; this happens because the weight of most particles tends to become negligible.
In other words most computational resources are wasted to explore low--probability regions of the state space.
To avoid this phenomenon, said \textit{weight degeneracy}, at each iteration the ESS is computed and if it is lower then a
threshold (set to $I/2$ in the simulations below) \textit{systematic resampling} is performed \cite{dojo11}, a procedure
that consists in removing low--weight particles and replacing them with copies of the high--weight ones: the resampled
set, with uniform weights, is an alternative representation of the same distribution.

\section{A Semi--Analytic Approach for Conditionally Linear Models}\label{par:SMC_clm}

Consider now an inverse problem in which the noise is additive and Gaussian and the forward model has a linear-Gaussian
substructure. Our aim is to show how the algorithm presented in the previous section can be modified in order to exploit
these structures.

Suppose indeed that the unknown can be split into two sets of variables $X = (R, Q)$ so that
\begin{equation}\label{eq:sa_model}
Y = G(R)Q + E;
\end{equation}
observe that the following factorizations for the prior and posterior distributions hold:
\begin{equation}
\pi(x) = \pi(r,q) = \pi(r)\ \pi(q|r)
\end{equation}
\begin{equation}
\pi(x|y) = \pi(r,q|y) = \pi(r|y)\pi(q|r,y).
\end{equation}
Moreover suppose that
\begin{itemize}
\item[i)] $Q, R$ and $E$ are mutually independent,
\item[ii)] $\pi(q|r) = \mathcal{N} \left( q; \bar{q}_0, \Gamma_{q_0} \right)$,
\item[iii)] $\pi_{noise}(e) = \mathcal{N} \left( e; \bar{e}, \Gamma_{e} \right)$,
\end{itemize}
where $\mathcal{N} \left(z; m, \Gamma \right)$ is the Gaussian density of mean $m$ and covariance matrix $\Gamma$ evaluated in $z$.

Under these assumptions, for each realization $R=r$ the marginal likelihood $\pi(y|r)$ and the conditional posterior $\pi(q|r, y)$
can be computed analytically. Indeed $\pi(y|r)$ results to be a Gaussian distribution with mean
\begin{equation}\label{eq:m_marg_like}
G(r)\bar{q}_0 + \bar{e}
\end{equation}
and variance
\begin{equation}\label{eq:G_marg_like}
G(r)\Gamma_{q_0}G(r)^{T} + \Gamma_{e}.
\end{equation}
Also $\pi(q|y,r)$ is a Gaussian density with mean
\begin{equation}\label{eq:m_marg_post}
\bar{q}_0 + \Gamma_{q_0}G(r)^T \Gamma(r) ^{-1} (y - G(r)\bar{q}_0 - \bar{e})
\end{equation}
and variance
\begin{equation}\label{eq:G_marg_post}
\Gamma_{q_0} - \Gamma_{q_0}G(r)^T \Gamma(r) ^{-1} G(r) \Gamma_{q_0},
\end{equation}
where
\begin{equation}
\Gamma(r) := G(r)\Gamma_{q_0}G(r)^{T} + \Gamma_{e}.
\end{equation}
All these things considered the posterior $\pi(x|y)$ can be approximated through the following two--step algorithm, we refer to as semi--analytic SMC sampler.
\begin{description}
  \item[First step:] we approximate $\pi(r|y)$ through an adaptive SMC sampler.\\
   Reproducing Eq. (\ref{eq:pdfs_wholeSMC}) we sample sequentially from the distributions
 \begin{equation}\label{eq:pdfs_semi-analyticSMC}
 \pi_n^{SA}(r|y) = \frac{\pi(r)\pi(y|r)^{\alpha_n}}{{\pi^{SA}_n}(y)}
 \end{equation}
 where the sequence of exponents $0 = \alpha_1 < \dots < \alpha_N = 1$  is chosen adaptively, and the superscript $SA$ is introduced
 to distinguish these distributions from the marginals of the sequence defined in Eq. (\ref{eq:pdfs_wholeSMC}); the difference between these
 two sequences will be explained more in detail below.
  \item[Second step:] for each particle $R_N^i$ obtained at the last iteration of the SMC procedure for $\pi(r|y)$ we analytically
  compute $\pi(q|R_N^i, y)$ through Eq. (\ref{eq:m_marg_post}) and (\ref{eq:G_marg_post}).
\end{description}
The main advantage of this method with respect to a SMC sampler procedure applied to approximate the full posterior
$\pi(x|y)$ is that only the non--linear variables are sampled while we deal analytically with the linear ones. As we
will show by means of simulated experiments this fact leads to some statistical, in terms of a reduction of the Monte
Carlo variance of the estimators, and computational improvements.\\

\noindent
\textbf{Remark.} As already suggested, the distributions used in the first step of this semi--analytic approach, Eq.(\ref{eq:pdfs_semi-analyticSMC}), are not the marginal
distributions of the ones used in the full SMC sampler, Eq.(\ref{eq:pdfs_wholeSMC}). \\
Indeed marginalizing Eq. (\ref{eq:pdfs_wholeSMC}) we obtain
\begin{equation}\label{eq:comparison_marg1}
\pi_n(r|y) = \int{\pi_n(r,q|y)\ dq} = \frac{\pi(r)}{\pi_n(y)} \ \int{\pi(q|r)\pi(y|r,q)^{\alpha_n} dq},
\end{equation}
while from Eq.  (\ref{eq:pdfs_semi-analyticSMC})
\begin{equation}\label{eq:comparison_marg2}
\pi_n^{SA}(r|y) = \frac{\pi(r)}{\pi_n^{SA}(y)} \bigg(\int{\pi(q|r)\pi(y|r,q)\ dq} \bigg)^{\alpha_n}.
\end{equation}
Furthermore under assumption ii) and iii) that ensure a Gaussian distribution for the conditional prior $\pi(q|r)$
and for the likelihood function $\pi(y|r,q)$ we can obtain more explicit expressions that are respectively
\begin{equation}\label{eq:comparison_marg1b}
    \pi_n(r|y) = \frac{\pi(r)}{\widehat{\pi_n}(y)} \;  \mathcal{N}(y \; ; \; G(r)\bar{q}_0 + \bar{e}  \; ,  \; G(r)\Gamma_{q_0}G(r)^{T} + \frac{1}{\alpha_n} \Gamma_{e})
\end{equation}
\begin{equation}\label{eq:comparison_marg2b}
    \pi_n^{SA}(r|y) =  \frac{\pi(r) \det\left(\Gamma(r) \right)^{\frac{1-\alpha_n}{2}}}{\widehat{\pi_n}^{SA}(y)} \;
    \mathcal{N}(y \; ; \; G(r)\bar{q}_0 + \bar{e} \; ,  \; G(r)\frac{1}{\alpha_n} \Gamma_{q_0}G(r)^{T} + \frac{1}{\alpha_n} \Gamma_{e}),
\end{equation}
where
\begin{displaymath}
\widehat{\pi_n}(y) = \pi_n(y) \; \alpha_n^{\frac{N}{2}} \; \left[(2\pi)^{N} \; \det{\Gamma_e} \right]^{\frac{\alpha_n-1}{2}}
\ \ \ \
\widehat{\pi_n}^{SA}(y) = \pi_n^{SA}(y) \; \alpha_n^{\frac{N}{2}} \; (2\pi)^{\frac{N (\alpha_n-1)}{2}},
\end{displaymath}
being $N$ the size of the square matrix $\Gamma_e$, are the normalizing constants.\\

Coherently with the interpretation given in \cite{soluar14}, Eq. (\ref{eq:comparison_marg1b}) shows
that the distributions $\pi_n(r|y)$ differ from each other only for the noise covariance matrix which is multiplied
by $1 / \alpha_n$. Instead Eq. (\ref{eq:comparison_marg2b}) shows that the distributions used by the semi--analytic SMC sampler,
$\pi_n^{SA}(r|y)$, can be interpreted as the marginals of the posterior distributions of a Bayesian model in which (i)
the noise covariance matrix is multiplied by $1 / \alpha_n$, (ii) the covariance matrix of the prior for $q$ is also multiplied
by $1 / \alpha_n$, and (iii) the marginal prior $\pi(r)$ is multiplied by $\left( \det(\Gamma(r))\right)^{\frac{1-\alpha_n}{2}}$.

\section{Application to Magnetoencephalography}\label{par:MEG}

We present here the application of the semi--analytic approach described above to the analysis of magnetoencephalographic
data. The present study improves and extends the work presented in \cite{soluar14}, where a full SMC sampler was
used to approximate the full posterior $\pi(x|y)$. The improvement stems from the fact that the linear variables are marginalized,
and therefore are not sampled; as a consequence, it becomes computationally feasible to estimate the non--linear parameters using a whole
time--series rather than a single topography.

\subsection{The MEG inverse problem}

MEG is a non-invasive functional neuroimaging technique that records the weak magnetic field produced by
neural currents by means of an helmet--shaped array of SQUID sensors \cite{haetal93}.
The physical process that goes from the neural currents to the measured field is modelled by the
Biot-Savart equation \cite{sa87}. From the mathematical point of view, the reconstruction of neural currents from MEG
data is an ill-posed problem; indeed the Biot-Savart operator is compact and the solution is not unique
\cite{sa87,fokuma04,dafoka05}.
Since MEG data are recorded at a typical frequency of around 1,000 Hertz, estimation of neural currents can benefit
from the use of spatio--temporal analysis.
In this paper we use a multi--dipole model for the neural currents, i.e. neural currents are represented as the
superposition of a small number of point-like currents, termed currents dipoles, each of which models the activity of a small brain area.
A current dipole is an applied vector, described by a location and a dipole moment, containing strength and orientation of the
current. In this model, it is rather typical to assume that the dipole locations remain fixed for relatively long time intervals, from ten
to few hundred milliseconds, as they represent the activity of a given neural population. On the other hand, the dipole moments
change in time, as the number of firing neurons and the degree of synchronization between them changes.
As we will show, using this model the MEG inverse problem can be formalized by an equation of type (\ref{eq:sa_model})
and thus the semi--analytic approach can be used.\\

In order to implement the multi--dipolar approximation, first of all we discretize the brain volume into $N_C$ cells,
and for each cell we choose a reference point of position $\textbf{z}(c)$, $c = 1, \dots, N_C$. Then neural currents are modeled as the superposition
of a small number of current dipoles that are allowed to be located only in the points of the brain grid previously defined:
\begin{equation}\label{eq:multi_dipolar_model}
x = \sum_{k=1}^{d} q^{(k)} \delta(\textbf{z}', \textbf{z}(c^{(k)}))
\end{equation}
where $\delta(\cdot, \cdot)$ is the Kronecker delta, $d$ is the number of active dipoles, $c^{(k)}\ \in\ \Bbb{N},\ k = 1, \dots, d$ are the indices of the grid points where there is a dipole
and $q^{(k)}\ \in\ \Bbb{R}^3,\ k = 1, \dots d$ are applied vectors representing the moment of each dipole.

Thus in the framework of multi--dipolar approximation the MEG inverse problem consists in the estimation of the number of active dipoles,
their locations and their time--varying moments from the recorded magnetic field.
From the Biot-Savart equation we have that the magnetic field produced by neural currents (\ref{eq:multi_dipolar_model})
depends non--linearly on the number of sources and on their locations, and depends linearly on the dipole moments so that we can write

\begin{equation}
\textbf{y} = \widetilde{G}(r)\textbf{q} + \textbf{e}
\end{equation}
where
\begin{itemize}
\item $\textbf{y} = \left(y_1, \dots,y_{N_t} \right)$ is a vector of length $N_s \cdot N_t$, containing the recordings made by the
$N_s$ sensors for all the $N_t$ time points;
\item $r = (d,c^{(1)}, \cdots, c^{(d)})$ is the collection of the non--linear variables, i.e. the number of sources $d$ and the indices of the source locations
in the brain grid $c^{(k)}$;
\item $\widetilde{G}(r)$ is the block--diagonal matrix which is obtained as
\begin{equation}
\widetilde{G}(r) = \left( \begin{array}{cccc} G(r) & 0  & \cdots & 0 \\
                                   0 &  G(r) & \ddots & \vdots \\
                                   \vdots & \ddots & \ddots & 0 \\
                                   0 & \cdots & 0 & G(r) \end{array} \right).
\end{equation}
where $G(r)$ is a matrix of size $(N_s) \times (3\cdot d)$; each column of $G(r)$ contains the magnetic field produced by a unit dipole,
placed in one of the grid points contained in $r$, and oriented along one of the three orthogonal directions; the number of blocks is equal
to the number of time--points, in accordance with the assumption that the number of sources and their locations do not change with time;
$\widetilde{G}(r)$ is therefore of size $(N_s \cdot N_t) \times (3 \cdot d \cdot N_t)$;

\item $\textbf{q} = \left(q_1^{(1)}, \dots ,q_1^{(d)}, \dots, q_{N_t}^{(1)}, \dots,q_{N_t}^{(d)} \right) $ is a vector of length $ 3 \cdot d \cdot N_t$, containing
the dipole moments of the $d$ dipoles at all $N_t$ time points;
\item $\textbf{e} = \left(e_1 \dots, e_{N_t}\right)$ is a vector of length $N_s \cdot N_t$, containing the noise affecting the measurements.
\end{itemize}

Therefore, under suitable Gaussian assumptions about the noise model and the prior for the dipole moment we can use the
semi--analytic approach.
As we will show in the next section, the computation can be simplified if we assume the independence between different time points,
that means we assume a block--diagonal matrix as the covariance matrix for the Gaussian distribution of noise and prior.

\subsection{Statistical model and algorithm settings}

In order to apply the semi--analytic approach described in the previous section, we need to define the Bayesian model,
i.e. the prior distribution and the likelihood function, and the transition kernels $K_k(r_{k-1}, r_k)$ used for
the evolution of the particles in the SMC procedure. \\

The parameters of the prior distribution are set in order to reflect neurophysiological knowledge and to
satisfy the constraints imposed by the semi--analytic approach.
For the distributions over the number of dipoles and their location we follow the same strategy proposed in
\cite{soluar14}, that we report here in the interest of completeness. The prior density for the dipole moment is
taken to be a Gaussian distribution in order to fulfill the requirements of the semi--analytic approach.

$\pi(r,\textbf{q})$ is decomposed into the product:
\begin{equation}
\begin{split}
\pi(r,\textbf{q}) & = \pi(d,c^{(1)}, \dots, c^{(d)}, q_1^{(1)}, \dots ,q_1^{(d)}, \dots, q_{N_t}^{(1)}, \dots,q_{N_t}^{(d)})\\
              & = \pi(d)\pi(c^{(1)}, \dots, c^{(d)}|d)\pi(q_1^{(1)}, \dots ,q_1^{(d)}, \dots, q_{N_t}^{(1)}, \dots,q_{N_t}^{(d)}|d, c^{(1)}, \dots, c^{(d)}).
\end{split}
\end{equation}
where $\pi(d)$ regards the number of dipoles and is chosen to be a Poisson distribution with small parameter;
for the dipole position we choose $\pi(c^{(1)}, \dots, c^{(d)}|d)$  to be a uniform distribution under the constraint
that different dipoles must occupy different positions. Finally we assume that
$\pi(q_1^{(1)}, \dots ,q_1^{(d)}, \dots, q_{N_t}^{(1)}, \dots,q_{N_t}^{(d)}|d, c^{(1)}, \dots, c^{(d)})$ is a zero--mean
Gaussian distribution whose covariance matrix $\Gamma_{\textbf{q}}$ is block--diagonal

\begin{equation}
\Gamma_\mathbf{q} = \left( \begin{array}{cccc} \Gamma_{q_1^{(1)}} & 0  & \cdots & 0 \\
                                   0 &  \Gamma_{q_1^{(2)}} & \ddots & \vdots \\
                                   \vdots & \ddots & \ddots & 0 \\
                                   0 & \cdots & 0 & \Gamma_{q_{N_t}^{(d)}} \end{array} \right), 
\end{equation}

where $\Gamma_{q_t^{(k)}}$ is the covariance matrix of the individual dipole, that may contain prior information
about the local source orientation. This choice corresponds to treating all the time points independently.
In the simulations below, we will be using $\Gamma_{q_t^{(k)}} = \sigma_q^2 \textbf{I}_{3d}$ and discuss the impact of the choice
of $\sigma_q$ on the results; here $\textbf{I}_{3d}$ is the identity matrix of order $3d$.\\

As far as the likelihood function is concerned, we choose a zero--mean Gaussian distribution; again, we assume that
we can treat the different time points independently, i.e. that noise has no temporal correlation, so that the covariance
matrix $\Gamma_{\textbf{e}}$ is block--diagonal

\begin{equation}
\Gamma_\mathbf{e} = \left( \begin{array}{cccc} \Gamma_{e} & 0  & \cdots & 0 \\
                                   0 &  \Gamma_{e} & \ddots & \vdots \\
                                   \vdots & \ddots & \ddots & 0 \\
                                   0 & \cdots & 0 & \Gamma_{e} \end{array} \right),
\end{equation}

where $\Gamma_e$ is the spatial covariance matrix; in the simulations below we will be using $\Gamma_e = \sigma_{e}^2 \textbf{I}_{N_s}$.

Under these assumptions the marginal likelihood $\pi(\textbf{y}|r)$ can be written as the product of the marginal
likelihoods for single time--points:
\begin{equation}\label{eq:complex_like}
\pi(\textbf{y}|r) = \prod_{t=1}^{N_{t}} \mathcal{N}(y_t; \textbf{0}, \Gamma_t(r))
\end{equation}
where $\Gamma_t(r) = G(r)\Gamma_{q_t}G(r)^T + \Gamma_e$, $\Gamma_{q_t}$ being the submatrix of $\Gamma_\mathbf{q} $
containing the covariance for the $d$ dipoles at a given time point, and each Gaussian is zero--mean because the
prior for the dipole moments and the likelihood are zero--mean. \\

Finally, the conditional posterior $\pi(\textbf{q}|r, \textbf{y})$ is a Gaussian density with mean
\footnotesize{
\begin{equation}
\left( \begin{array}{c}
\Gamma_{q_1}G(r)^T \Gamma_1(r) ^{-1} (y_1)\\
\vdots \\
\Gamma_{q_{N_t}}G(r)^T \Gamma_{N_t}(r) ^{-1} (y_{N_{t}}) \end{array} \right)
\end{equation}}
\normalsize
and variance
\footnotesize{
\begin{equation}
\left( \begin{array}{cccc}
\Gamma_{q_1} - \Gamma_{q_1}G(r)^T \Gamma_1(r) ^{-1} G(r) \Gamma_{q_1} & 0 & \cdots & 0 \\
0 & \Gamma_{q_2} - \Gamma_{q_2}G(r)^T \Gamma_2(r) ^{-1} G(r) \Gamma_{q_2} & \cdots & 0 \\
\vdots & \vdots & \vdots & \vdots \\
0 & \cdots & 0 &  \Gamma_{q_{N_t}} - \Gamma_{q_{N_t}}G(r)^T \Gamma_{N_t}(r) ^{-1} G(r) \Gamma_{q_{N_t}}
\end{array} \right).
\end{equation}}
\normalsize
Thus the distributions over the dipole moments associated to different instants can be treated separately.\\

In order to implement the SMC sampler, we have to build the transition kernels $K_k(r_{k-1}, r_k)$ that we assume to be $\pi_k(r_k|y)$ invariant.
Because the number of dipoles is unknown, we use a variable dimension model and thus we have to jump between spaces of different
dimensions, i.e. the state spaces of sets with different number of dipoles. \\
As suggested in \cite{soluar14}, to do this we split the evolution into two steps. First we treat the evolution of the
number of dipoles through a Reversible Jump Metropolis-Hastings \cite{gr95} that accounts for a possible birth or
death move. More specifically the proposal density is built as follows. The birth of a new dipole is proposed with
probability $P_{birth} = \frac{1}{3}$ and the location of the new dipole is uniformly drawn from the brain grid points
not yet occupied; otherwise the death of a dipole, uniformly drawn from the ones that compose the particle, is proposed
with probability $P_{death} = \frac{1}{20}$. If no birth or death are proposed or if the suggested move is rejected
the particle doesn't change.\\
After we deal with the evolution of the location of all the dipoles that compose the particle, even the possibly new ones.
The evolution of each dipole is treated separately through a Metropolis-Hastings kernel. The new dipole location is
drawn from the grid points within a radius of $1$ cm from the old position with probability proportional to a Gaussian
centered at the starting point.\\

We end this Section with a brief description of a computational trick we need in order to implement the SMC procedure for
the marginal posterior $\pi(r|\textbf{y})$. As described in Section \ref{Sec:BIPSMC}, at each iteration $n$ the unnormalized weight
associated to the $i$--th particle is computed through the recursive formula
\begin{equation}
w_{n+1}^i = W_{n}^i \pi(\textbf{y}|R_n^i)^{\alpha_{n+1} - \alpha_{n}}
\end{equation}
thus the likelihood $\pi(\textbf{y}|R_n^i)$ has to be evaluated.\\
If the hypothesis of independence just described holds, $\pi(\textbf{y}|R_n^i)$  is equal to the product of the marginal likelihood over
the $N_t$ instants of the time window.\\
From a computational point of view that means we have to multiply $N_t$ factors that are lower than 1: this may cause underflow
for almost all values of $N_t$ except very low ones.\\
In order to prevent this fact we introduce the log--likelihood function and we proceed as follows:
\begin{enumerate}
\item for each particle we compute the logarithm of the unnormalized weight
   \begin{equation}
    \log(w_{n+1}^i) = \log(W_n^i) + (\alpha_{n+1}-\alpha_n)\sum_{t=1}^{N_t} \log(\pi(y_t|R_n^i));
    \end{equation}
\item then we compute the logarithm of the normalizing constant $C_{norm} := \sum_{i=1}^I w_{n+1}^i$ through the \textit{log-sum-exp formula}
    \begin{equation}
    \log(C_{norm}) = w + \log(\sum_{i=1}^I e^{\log(w_{n+1}^i)-w})
    \end{equation}
    where $w = \max_i \{\log(w_{n+1}^i)\}$;
\item finally we compute the logarithm of the normalized weight
    \begin{equation}\label{eq:log_norm_weight}
    \log(W_{n+1}^i) = \log(w_{n+1}^i) - \log(C_{norm})
    \end{equation}
    and then, only after we have normalized we calculate the exact value of the weights taking the exponential of Eq. (\ref{eq:log_norm_weight}).
\end{enumerate}
Observe that the same computational trick can be used to calculate the ESS from the logarithm of the normalized weights $\{W_{n+1}^i \}_{i=1}^I$.

\subsection{Point estimation}\label{par:point_est}

As described in Section \ref{Sec:BIPSMC}, at the end of the SMC procedure for $\pi(r|\textbf{y})$ we obtain a Monte Carlo approximation for the posterior itself
through the weighted particles $\{R^i, W^i\}_{i=1}^I$, where the index $N$ of the iteration is henceforward omitted for simplicity of notation.
From this approximation, point estimates for the number of active dipoles and their location can be obtained as described in \cite{soluar14};
then a point estimate for the dipole moments over time can be straightforwardly computed from the analytical expression of the conditional posterior $\pi(\textbf{q}|r, \textbf{y})$.
More specifically, the estimated number of sources $\widehat{D}$ is defined as the mode of the marginal distribution of the number of dipoles that can be calculated as
\begin{equation}
\mathbb{P}(D = d|\textbf{y}) = \sum_{i=1}^I W^i \delta(d,D^i)
\end{equation}
where $D^i$ is the number of dipoles of the $i-$th particle $R^i$.\\

The estimated locations of the $\widehat{D}$ active sources are the $\widehat{D}$ highest local modes of the intensity measure for the source location
conditioned on the estimated number of sources, that can be computed as
\begin{equation}\label{est_intensity_measure}
\mathbb{P}(c|\textbf{y}, \widehat{D}) = \sum_{i=1}^I W^i \delta(\widehat{D}, D^i) \sum_{k=1}^{D^i} \delta(c,C^{(k),i})
\end{equation}
where $c$ is a point of the brain grid and $C^{(k),i}$, $ k = 1, \dots, D^i$, is the location of the $k$--th dipole into the $i$--th particle.
Observe that only the particles with the correct number of dipoles contribute to this measure.\\

Finally, an estimate of the time--varying dipole moments
$\left( \widehat{Q}^{(1)}_1, \dots \widehat{Q}^{(\widehat{D})}_1,
\dots \widehat{Q}^{(1)}_{N_t}, \dots \widehat{Q}^{(\widehat{D})}_{N_t}  \right)$
is obtained analytically, as the mean of the conditional distribution
\begin{equation}\label{eq:pdf_est_dipole}
\pi(q^{(1)}_1, \dots, q^{(\widehat{D})}_1, \dots, q^{(1)}_{N_t}, \dots, q^{(\widehat{D})}_{N_t}
|\textbf{y}, \widehat{D}, \widehat{C}^{(1)}, \dots \widehat{C}^{(\widehat{D})})
\end{equation}
that, in accordance with the results proved in the previous sections, is a Gaussian density.

\section{Simulation Experiments}\label{par:Simulation}
In this section simulated data are used to validate and assess the performance of the semi--analytic approach.
More specifically in Section \ref{par:exp2} we use a large number of datasets to investigate the behaviour of the
algorithm under various experimental conditions, as well as the impact of a partially wrong prior.
In Section \ref{par:exp1} a comparison between our approach and the full SMC presented in \cite{soluar14} is made
in terms of the Monte Carlo variance of the approximation obtained. Finally in Section \ref{par:exp3} we present
some computational considerations about the computational cost of the algorithm.

\subsection{Experiment 1: Validation of the Method}\label{par:exp2}

Following \cite{soetal13,soluar14} we quantify the performances of the method by calculating discrepancy measures between
the true and the estimated source configuration. Let $(D, C^{(1:D)}, Q^{(1:D)}_1, \dots, Q^{(1:D)}_{N_t})$ and
$(\widehat{D}, \widehat{C}^{(1:\hat{D})}, \widehat{Q}^{(1:\hat{D})}_1, \dots, \widehat{Q}^{(1:\hat{D})}_{N_t})$ be the true and the estimated dipole configuration, respectively;
we use:
\begin{itemize}
\item $\Delta_d$ which is the difference between the true and the estimated number of dipoles $\Delta_d = \widehat{D}-D$;
\item $\Delta_c$ that quantifies the localization error and is defined as
\begin{equation}
\Delta_c = \begin{cases}
            \min_{\lambda \in \Lambda_{\widehat{D}, D}}{\frac{1}{\widehat{D}} \sum_{k=1}^{\widehat{D}}{ |\ \textbf{z}(C^{(k)}) - \textbf{z}(\widehat{C}^{(\lambda(k))})\ |}}\ \mbox{if}\ D \geq \widehat{D}   \\
            \min_{\lambda \in \Lambda_{D, \widehat{D}}}{\frac{1}{D}  \sum_{k=1}^{D}{|\ \textbf{z}(C^{(\lambda(k))}) - \textbf{z}(\widehat{C}^{(k)})\ |}}\ \mbox{if}\ D < \widehat{D}
            \end{cases},
\end{equation}
where $\Lambda_{k, l}$ is the set of all the permutations of $k$ elements drawn from $l$ elements.\\
$\Delta_c$ is a modified version of the OSPA metric with no penalty for cardinality errors, which are evaluated separately by $\Delta_d$ above.
For more details see \cite{scvovo08}.
\end{itemize}
In addition, we compare the true and the estimated time courses, i.e. the norm of the dipole moment as a function of time.\\

In a first series of tests we generated 300 datasets, each one made of 30 time points and containing 2, 3 or 4 sources.
For each dataset, dipole locations are uniformly drawn from the brain grid points with a reciprocal distance of at least 1 cm.
Dipole orientations are uniformly drawn from the unit sphere and do not change in time.
In half datasets the source time courses are independent: dipoles are active one after the other, with almost negligible
temporal overlap; in the other 150 datasets, the sources have exactly the same time course. In the MEG literature this last
condition is usually defined as sources being perfectly \textit{correlated}; while such condition often happens in real scenarios,
many well--known inverse methods, such as MUSIC \cite{molele92, mole99} and beamformers \cite{vvetal97,seetal02}, encounter difficulties
in estimating perfectly correlated sources. In conclusion, in this first experiment the datasets are divided in six groups, according to
the number of sources and their temporal correlation.

\begin{figure} [h!]
    \centering
    \subfigure[High SNR]{\includegraphics[scale=0.5]{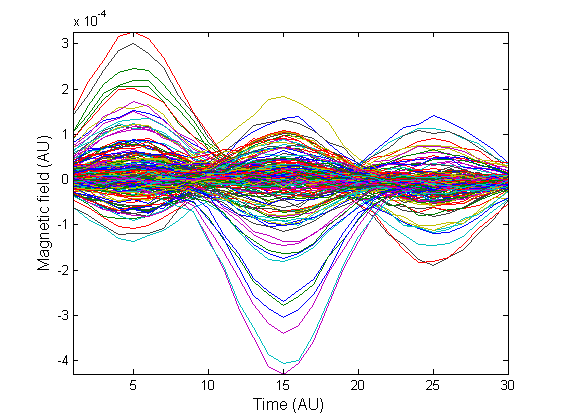}}\ \
    \subfigure[Low SNR]{\includegraphics[scale=0.5]{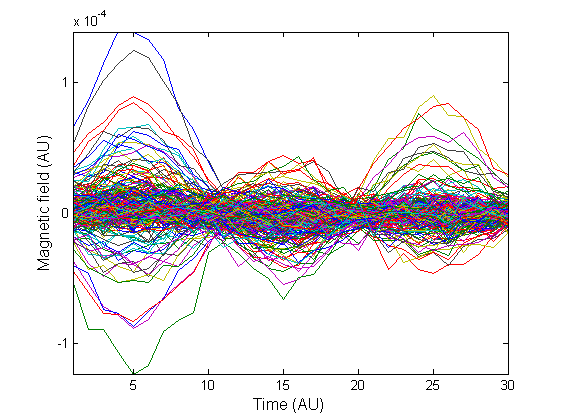}} \\
    \caption{Two synthetic MEG time--series generated by 3 uncorrelated dipoles. While all the datasets have been perturbed with noise of the same standard deviation,
    the resulting signal--to--noise ratio varies considerably due to the different locations and orientations of the sources, that produce signals of different
    intensities.}\label{fig:es_magnetic_field}
\end{figure}

Synthetically generated data were then perturbed with white Gaussian noise of fixed standard deviation; because the source
location and orientation are random, the actual signal--to--noise (SNR) ratio varies considerably for different sources.
As an example Figure \ref{fig:es_magnetic_field} shows two synthetic MEG times series, both generated by 3 uncorrelated dipoles
but with different SNR.

We analyze the 300 datasets setting $I = 1000$, that is a good compromise between the quality of the results and the computational cost;
the value $\sigma_{e}$ in the likelihood is set equal to the standard deviation of noise, while the parameter $\sigma_{q_0}$ for the prior over the
dipole moment is set equal to 1, according to the dipole strength we have simulated.

\begin{figure} [h!]
    \centering
    \subfigure[2 uncorrelated dipoles]{\includegraphics[scale=0.3]{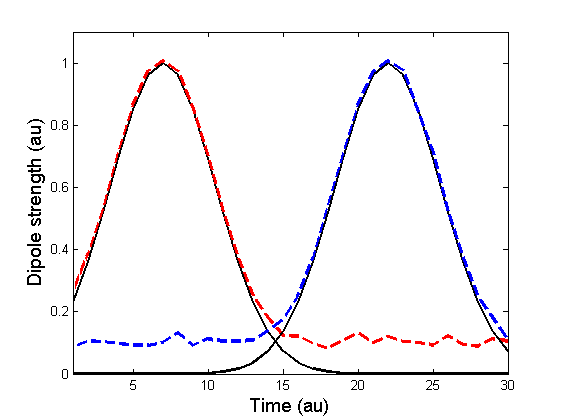}}\ \
    \subfigure[3 uncorrelated dipoles]{\includegraphics[scale=0.3]{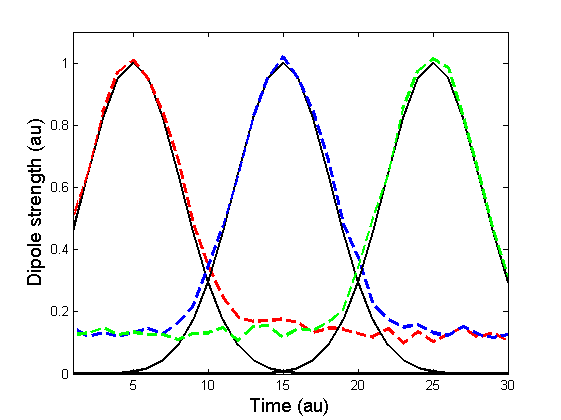}}\ \
    \subfigure[4 uncorrelated dipoles]{\includegraphics[scale=0.3]{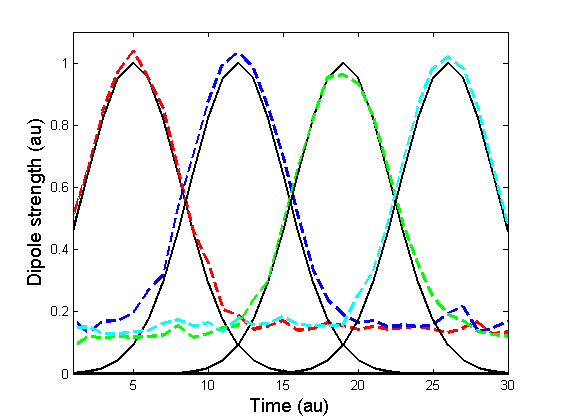}}\ \
    \subfigure[2 correlated dipoles]{
    \includegraphics[scale=0.25]{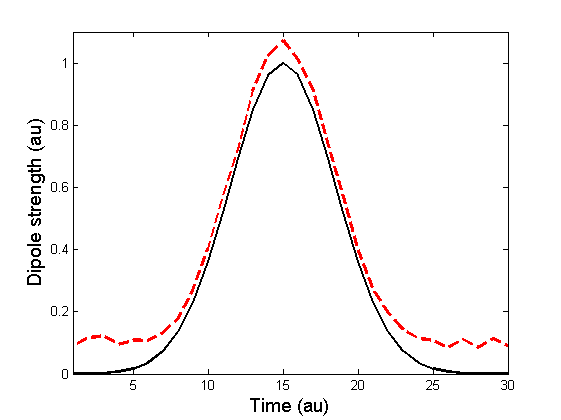}\
    \includegraphics[scale=0.25]{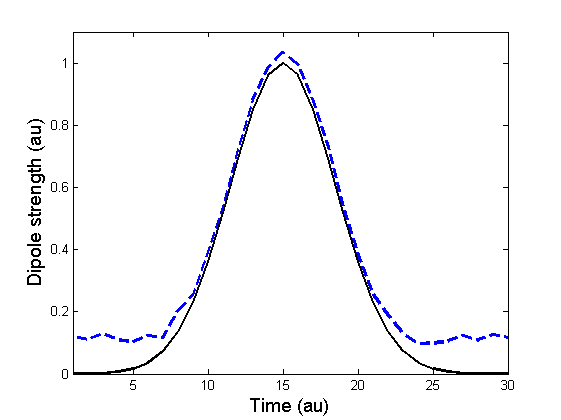}}
    \subfigure[3 correlated dipoles]{
    \includegraphics[scale=0.25]{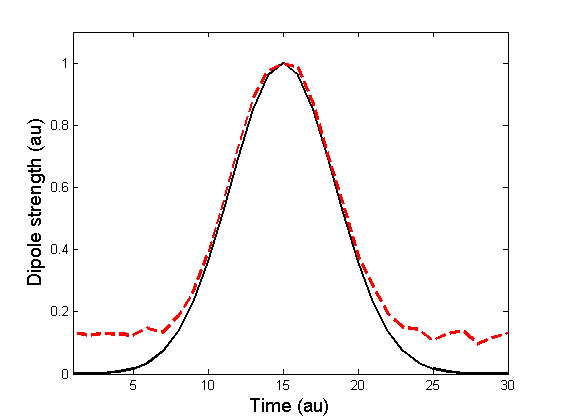}\
    \includegraphics[scale=0.25]{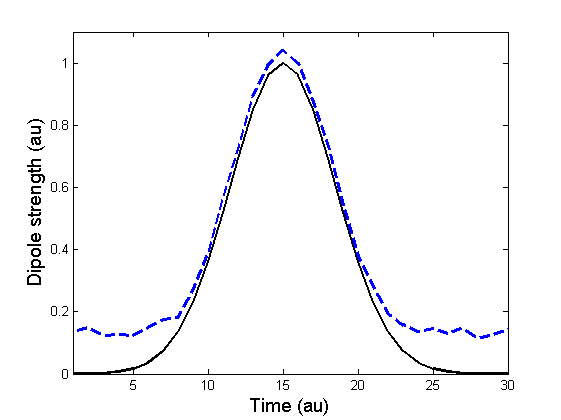}\
    \includegraphics[scale=0.25]{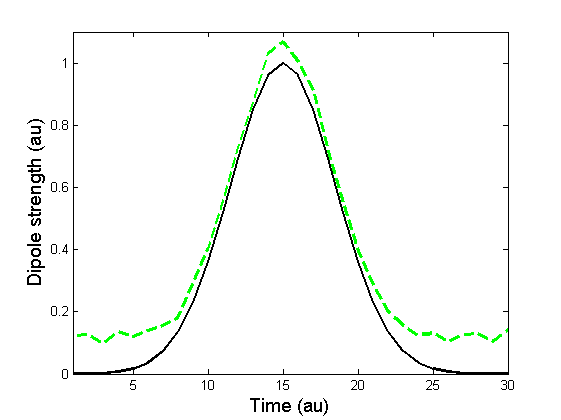}}
    \subfigure[4 correlated dipoles]{
    \includegraphics[scale=0.25]{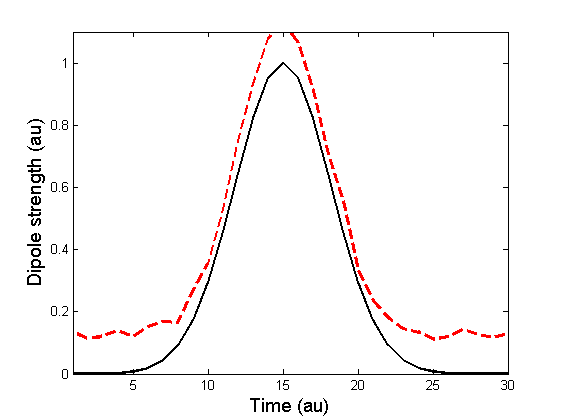}\
    \includegraphics[scale=0.25]{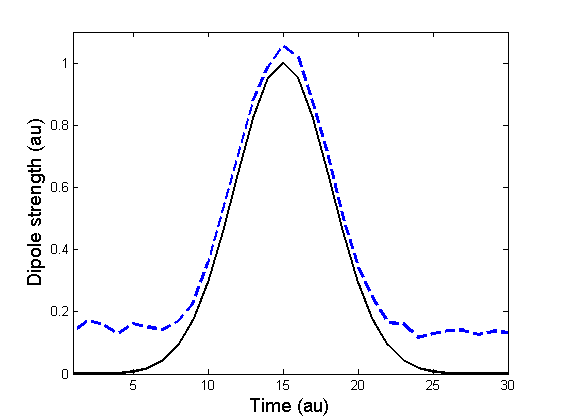}\
    \includegraphics[scale=0.25]{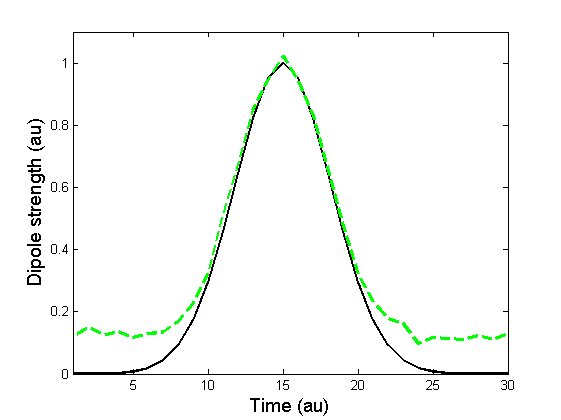}\
    \includegraphics[scale=0.25]{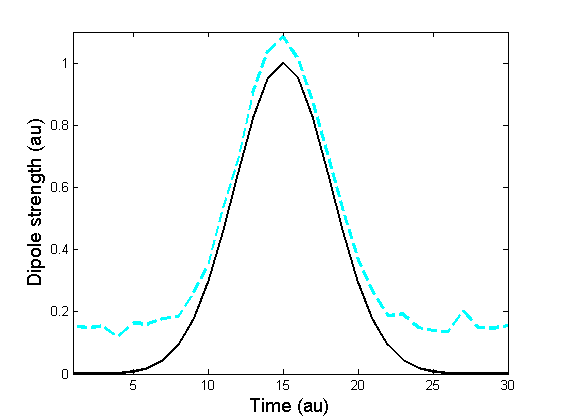} }
    \caption{Reconstructed dipole strength, $\lVert Q^{(k)}_t \rVert$, averaged over runs (dotted lines) superimposed on the true ones (black solid line).}\label{fig:error_strenght}
\end{figure}

Table \ref{tab:delta} shows the values of the discrepancy measures $\Delta_d$ and $\Delta_c$ averaged over runs together with their
standard deviations. Figure \ref{fig:error_strenght} shows the true and the estimated source time courses, again averaged over runs.

\begin{table} [h!]
    \centering
    \subtable{
    \begin{tabular}{|c||c|c|}
    \hline
    Unc. & $\Delta_d$ & $\Delta_c$\\
    \hline
    \hline
    2 dip &  0.00 $\pm$ 0.00 & (0.3 $\pm$ 1.1) mm \\
    3 dip & -0.04 $\pm$ 0.19 & (0.9 $\pm$ 1.9) mm \\
    4 dip & -0.04 $\pm$ 0.20 & (1.0 $\pm$ 2.0) mm \\
    \hline
    \end{tabular}} \hspace{-.2cm}
    \subtable{
    \begin{tabular}{|c||c|c|}
    \hline
    Corr. & $\Delta_d$ & $\Delta_c$\\
    \hline
    \hline
    2 dip &  0.00 $\pm$ 0.00 & (0.7 $\pm$ 2.6) mm \\
    3 dip & -0.08 $\pm$ 0.27 & (1.0 $\pm$ 2.6) mm \\
    4 dip & -0.12 $\pm$ 0.33 & (1.1 $\pm$ 1.9) mm \\
    \hline
    \end{tabular}}
    \caption{Discrepancy measures for the number of dipoles (left) and their location (right) averaged over 50 runs
    for different numbers of active sources and different levels of correlation.}\label{tab:delta}
\end{table}

There seems to be a very small difference between the correlated and the uncorrelated case, as far as the estimation of the number
of dipoles and of the dipole locations are concerned: the discrepancy measures are slightly higher for the correlated case,
where the number of dipoles is under--estimated more ($\Delta_d$ is lower then zero) and the  average localization error
is slightly higher. These two things are of course related: when two sources are very close
to each other, it sometimes happens that the algorithm estimates a single source in between, thus contributing to increasing
both discrepancies. On the other hand, both discrepancy measures $\Delta_d$ and $\Delta_r$ tend to increase with the number of dipoles.
This is of course expected, because, as the number of dipoles increases: (i) the ill--posedeness of the problem gets worse, i.e. there may be more alternative
configurations explaining the data equally well; (ii) the state--space to be explored increases dramatically, and therefore it becomes more
likely to miss the high--probability region; (iii) given the random generation of the data, it becomes more likely that two sources happen to be
in nearby locations, and can therefore be explained by a single source. Indeed, a similar behaviour was also observed in \cite{soluar14}.\\
Figure \ref{fig:error_strenght} conveys additional information about the proposed method. First, we observe that when the
true source strength is zero, the estimated source strength is not (on average). This is in fact a well known pitfall of multi--dipole
models with a fixed number of sources: in the temporal window where an estimated source is not actually active, its estimated strength is
tuned to optimally explain the noise component in the data.
On the other hand, in the case of correlated dipoles the source strengths appear to be slightly over--estimated also at the peak.
Going through the datasets one by one, we have observed that this is not a systematic over--estimation, but rather the consequence of occasional and diverse estimation errors in a few datasets: in two two--dipole cases, the estimated source locations were deeper than the true ones,
thus requiring stronger dipoles to reproduce the measured field; in a couple of three--dipole and four--dipole cases, under--estimation of the number of
sources, due to the proximity of two dipoles, led to over--estimation of the strength.\\

In a second series of tests, we investigated what is the effect of a mis--specification of the prior distribution; namely, we wanted
to investigate the robustness with respect to the parameter $\sigma_q$, that is the standard deviation of the prior for the dipole moment.
To do this, we produced 100 datasets, each one generated by two dipoles of random location; the first dipole has a peak strength of 1, the
second dipole has a peak strength of 10; we then applied the algorithm with three different values of $\sigma_q$: $1$, $5$ and $10$.
In Table \ref{tab:sigmaq} we report the discrepancy measures averaged over these 100 datasets. The results indicate that a mis--specification
of the prior can lead to the following cases.
When the standard deviation of the prior is 5 or 10, the number of dipoles tends to be under--estimated, because it becomes
more likely to miss the weak source.
When the standard deviation of the prior is $1$, on the other hand, the number of sources tends to be over--estimated, because
it sometimes happens that the strong dipole is replaced by two close--by dipoles, whose strengths summing up to 10 (the strength of the true source).
While relatively unpleasant, this is somehow expected, due to the fact that a Gaussian distribution
gives extremely low probability outside of the $\pm 3 \; \sigma$ range; therefore, despite the Poisson prior discouraging larger models, a two--dipole
model with small intensities has higher prior probability than a single--dipole model with strong dipole moment; since the likelihood of the two configurations is the same,
the posterior peaks on the two--dipole model.

\begin{table} [h!]
    \centering
    \begin{tabular}{|c||c|c|}
    \hline
    & $\Delta_d$ & $\Delta_c$ \\
    \hline
    \hline
    $\sigma_q = 1$ & 0.20 $\pm$ 0.71  & (1.96 $\pm$ 3.65) mm \\
    $\sigma_q = 5$ & -0.20 $\pm$ 0.40 & (2.53 $\pm$ 9.71) mm \\
    $\sigma_q = 10$& -0.26 $\pm$ 0.44 & (1.41 $\pm$ 3.99) mm \\
    \hline
    \end{tabular}
    \caption{Discrepancy measures for the number of dipoles (left) and their location (right) averaged over 100 runs
    for different values of the parameter $\sigma_q$.}
    \label{tab:sigmaq}
\end{table}

\subsection{Experiment 2: Variance Comparison}\label{par:exp1}
In order to investigate the differences between the semi--analytic approach and the full SMC presented in \cite{soluar14} we produced
a dataset containing only a single time point; the data are generated by two sources, one deeper than the other.
Dipole strenghts were both set to 1 while their directions are along one of the coordinate axes. White Gaussian noise was added,
with standard deviation set to $5 \%$ of the peak of the noise-free signal, corresponding to a SNR similar to
that of evoked responses.

We do 200 runs of both the algorithms, the semi--analytic SMC and the full SMC, with the same data as input. We set the number
of particles $I$ to 100, we use a Gaussian prior for the dipole moment with zero mean and covariance matrix $\Gamma_{q_0} = \textbf{I}$ and a
Gaussian likelihood with zero mean and covariance matrix $\Gamma_{e} = \sigma_{e}^2 \textbf{I}$ where $\sigma_{e}$ is set to the true value,
i.e. the value actually used to simulate the noise.

\begin{figure} [h!]
\centering
\includegraphics[scale=0.7]{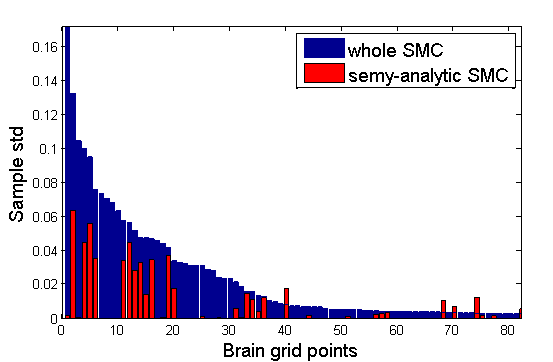}
\caption{Sample standard deviation over the brain grid point for the two algorithms.}\label{fig:variance_comparison}
\end{figure}

\begin{figure} [h!]
    \centering
    \subfigure{\includegraphics[scale=0.45]{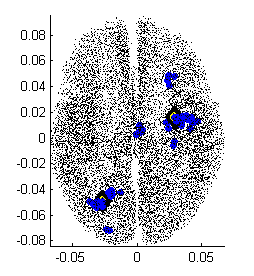}}\ \
    \subfigure{\includegraphics[scale=0.45]{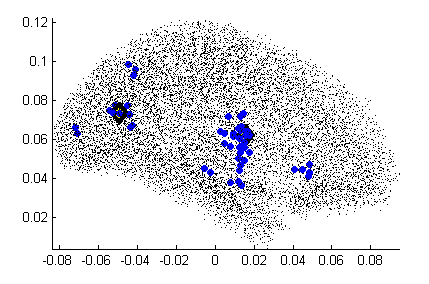}} \\
     \subfigure{\includegraphics[scale=0.45]{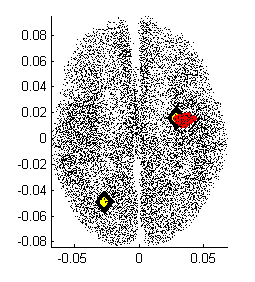}}\ \
    \subfigure{\includegraphics[scale=0.45]{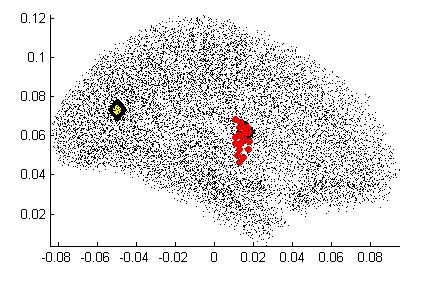}}
    \caption{Sample standard deviation on the brain. In each figure we plot the true location of sources (yellow diamond), and
     the brain grid points whose standard deviation is over a threshold equal to $5\cdot10^{-3}$ (blue points for the full SMC,
    first raw, and red points for the semi--analytic SMC, second raw).}\label{fig:variance_position} 
\end{figure}

In order to quantify the variance between different runs we calculate the sample standard deviation of the
intensity measure for the source location in the following way. For each run $l$, we compute the quantity
\begin{equation}
\mathbb{P}_l (c|y) \simeq  \sum_{i=1}^I  W^i \sum_{k=1}^{D^i} \delta(c, C^{(k),i}).
\end{equation}
that is an approximation to the intensity measure, differing from equation (\ref{est_intensity_measure}) because here there is no conditioning
on the estimated number of dipoles; therefore, all particles contribute to the calculation of this quantity.
Then we compute the sample standard deviation of $\mathbb{P}_l (c|y) $ over runs. In Figure \ref{fig:variance_comparison} we compare the
standard deviations produced by the two algorithms, the full SMC (dark blue bar)
and the semi--analytic SMC (red bar), on the subset of points where such standard deviation is higher than $5 \cdot 10^{-3}$.
The values of the standard deviation of the full SMC are sorted in descending order, then the standard deviations of the semi--analytic
SMC on the corresponding brain grid points are superimposed: the semi--analytic SMC clearly produces lower--variance approximations
of the posterior distribution. In Figure \ref{fig:variance_position} we provide the complementary information of what grid points
are affected by higher variance: as expected, these points concentrate around the true sources (yellow diamonds), where the posterior probability is higher.
The Figure also demonstrates that the points involved by the semi--analytic SMC are fewer.
Finally, both algorithms produce higher variance in correspondence of the deeper source.


\subsection{Experiment 3: Dependence of the Execution--Time on the Length of the Time Window.}\label{par:exp3}

In this section we study how the run--time of the algorithm depends on the length of the time window given in input.
We do this by means of direct numerical simulation, because the computational cost of the algorithm is itself a Random Variable,
and can change considerably depending on the number of sources to be estimated, that, in turn, also tunes the adaptively chosen number of iterations.
In addition, the main practical advantage of the proposed method, against the full SMC, consists in opening the possibility to estimate dipoles from a whole
time window, maintaining, as we are going to show, the same computational cost needed for a single time point.

We proceed as follows. We consider one of the dataset described in the previous section in the case
of two correlated dipoles and we extract from it 5 time windows of different lengths, namely of length 1, 5, 10, 20, 30. All the time--series
are centered in the middle point of the original dataset ($t = 15$), to be sure that the true number of dipoles is the same in all the time windows.
Coherently with what we have done in the previous Section, the time--series are analyzed setting
$I = 1000$, $\sigma_{q_0} = 1$ and $\sigma_e$ equal to the standard deviation of the noise. In all the cases the number of dipoles and their
locations are estimated correctly and the amount of time spent by the algorithm is approximately the same, as Table \ref{tab:time1} shows.\\

\begin{table} [h!]
 \centering
 \begin{tabular}{|c||c|c|}
 \hline
 Window size & Time & Iteration \\
 \hline
 \hline
 1 &  54.01 min & 159\\
 5 &  56.97 min & 162\\
 10 & 50.22 min & 150\\
 20 & 46.68 min & 147\\
 30 & 50.19 min & 151\\
 \hline
 \end{tabular}\caption{Run--time in minutes, and number of iterations made when different time windows are given in input. The number of particles $I$ is set equal to 1000.}\label{tab:time1}
\end{table}

In order to understand this result observe that the vast majority of computational effort is spent in the evaluation of the marginal log--likelihood,
essential to calculate the particle weights. If the assumptions largely described in the previous Sections hold, when a time--window is given
in input the total marginal likelihood is the product of the likelihood at different time points; then, given a particle $R_n^i$
\begin{equation}\label{eq:factorization_likelihood}
\log(\pi(\textbf{y}|R_n^i)) = \sum_{t=1}^{N_t} \log(\pi(y_t|R_n^i)).
\end{equation}

Moreover, $\pi(y_t|R_n^i)$ is a zero--mean Gaussian distribution, whose covariance matrix
$\Gamma(R_n^i) = \sigma_{q_0}^2 G(R_n^i)G(R_n^i)^T + \sigma_e^2 \textbf{I}_{N_s}$ is the same at all time points.
Therefore
\begin{equation}
\log(\pi(\textbf{y}|R_n^i)) = -\frac{N_t}{2} \log\left(\det(\Gamma(R_n^i))\right) - \frac{1}{2}\sum_{t=1}^{N_t} \log \left(y_t^T \Gamma(R_n^i)^{-1}y_t \right) + c.
\end{equation}
where $c = - \frac{N_s \; N_t}{2} \; \log(2\pi)$ is an additive constant that is common to all particles and therefore does not have
to be actually calculated, because it disappears during the normalization.

Here, the most time--consuming operations are the computation of the determinant $\det(\Gamma(R_n^i))$
and of the inverse matrix $\Gamma(R_n^i)^{-1}$, of size $N_s \times N_s $; both operations
need to be done only once per particle, and do not depend on the length of the time window.
The only calculation that has to be repeated for all the time--points is the matrix-vector product $y_t^T \Gamma(R_n^i)^{-1}y_t$, which is negligible in terms of computational time.\\

Observe that this result is strictly due to the model structure, i.e., to the assumption of independence between different time points
and stationarity of the noise and prior covariances. Indeed, removing the assumption of independence, $\Gamma_{\textbf{q}}$ and/or $\Gamma_{\textbf{e}}$
would no longer be block--diagonal; as a consequence the factorization (\ref{eq:factorization_likelihood}) would not hold and
we would be forced to compute the high--dimensional Gaussian distribution $\pi(\textbf{y}|R_n^i)$.
In turn, this would require the computation of the determinant and the inversion of a large matrix,
of size $(N_s \cdot N_t) \times (N_s \cdot N_t)$. In this case, the computational cost would therefore depend on the length of
the time--window $N_t$ and may become potentially unbearable as $N_t$ increases.

\section{Application to Experimental Data}

\label{par:real_data}

We demonstrate the semi--analytic SMC
on an experimental dataset recorded during stimulation of the median nerve (Somatosensory Evoked Fields, SEF).
The somatosensory response to this type of stimuli is indeed relatively well understood and is often
used as a reference for validation on real data.
We use the very same dataset used in \cite{soluar14} and, as a further validation, we also report
here the source estimates obtained with the full SMC, and with two largely used inverse methods,
dynamic Statistical Parametric Mapping (dSPM \cite{daetal00}) and sLORETA \cite{pa02}.
Importantly, these three additional methods take in input a single time point rather than
a time--series; here we compare the estimates obtained by the semi--analytic SMC using a time window,
with those obtained by these methods at the signal peak in the same window.
In addition, the last two methods are not based on a dipolar source model, but
on a distributed source model instead: in particular, dSPM computes first the Tichonov solution
of the inverse problem, and then calculates a normalized version that turns out to be $t$-distributed under the null hypothesis of no activation.
SLORETA has a similar approach, but the resulting quantity has a different distribution.
Therefore, the comparison has to be considered as a qualitative comparison rather than a quantitative one.

We refer to \cite{soluar14} for a detailed description of the data and of the pre--processing steps.
Here we recall that the expected neural response to stimulation of the median nerve \cite{maetal97} comprises a first activation
in the primary somatosensory cortex around $25$ ms after the stimulus, in the hemisphere contralateral to the stimulation,
a later activation of parietal sources around $50$ ms after the stimulus, and finally possibly frontal sources around $100$ ms.

We applied the semi--analytic SMC with 1,000 particles. We used a diagonal covariance matrix in the likelihood function,
corresponding to assuming spatially uncorrelated noise; the standard deviation was estimated from the pre--stimulus interval,
by taking the maximum value among the standard deviations of individual channels.
The standard deviation of the prior for the dipole moment was set to 20 nAm. We analyzed the three time windows $17.5-35$ ms,
$37.5-75$ ms and $115-130$ ms.

\begin{figure}
\begin{center}

\includegraphics[width=3cm]{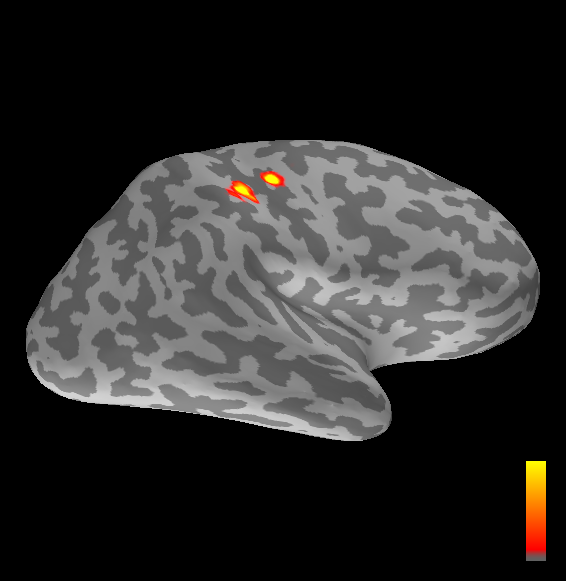}
\includegraphics[width=3cm]{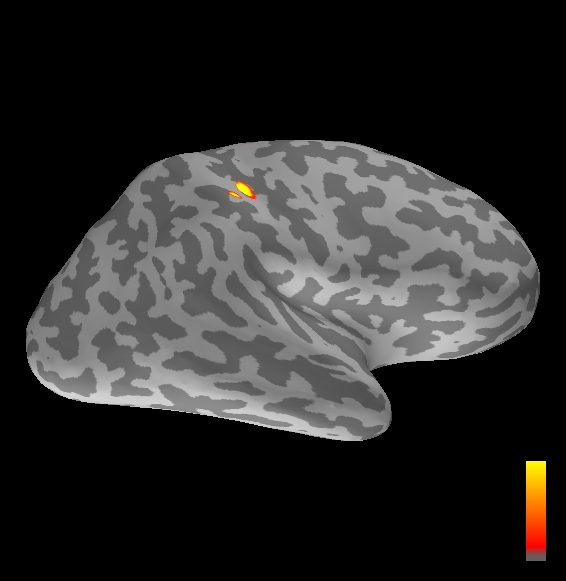}
\includegraphics[width=3cm]{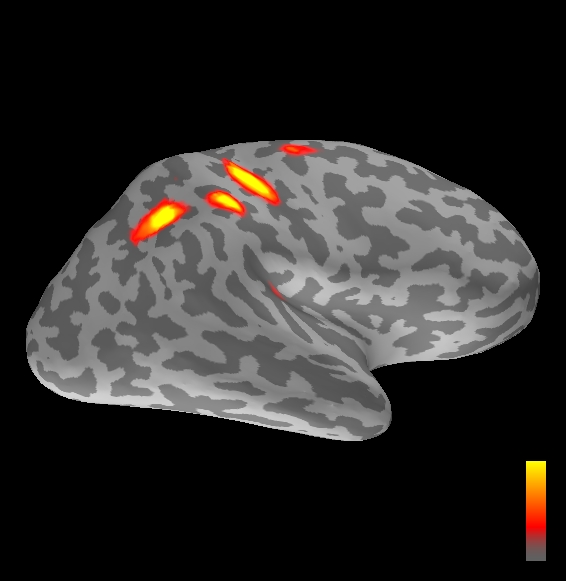}
\includegraphics[width=3cm]{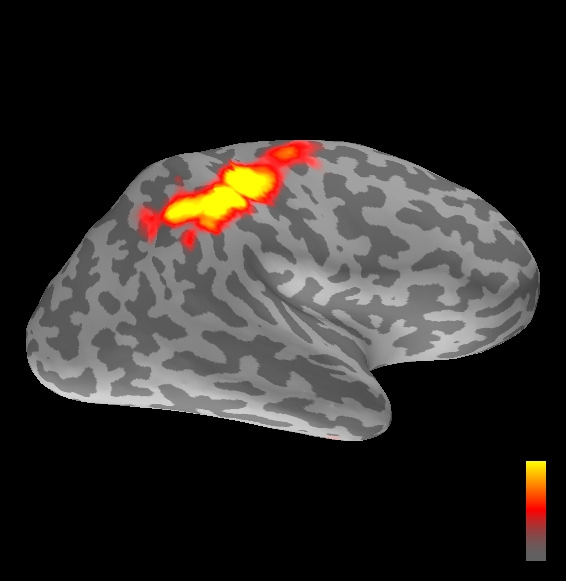}\\[4pt]
\includegraphics[width=3cm]{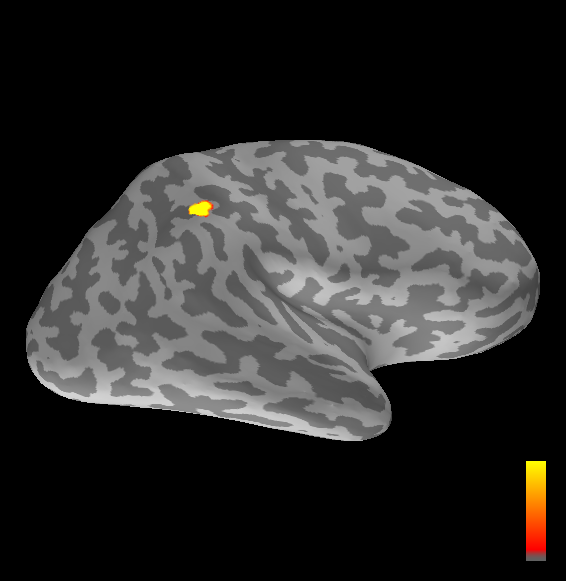}
\includegraphics[width=3cm]{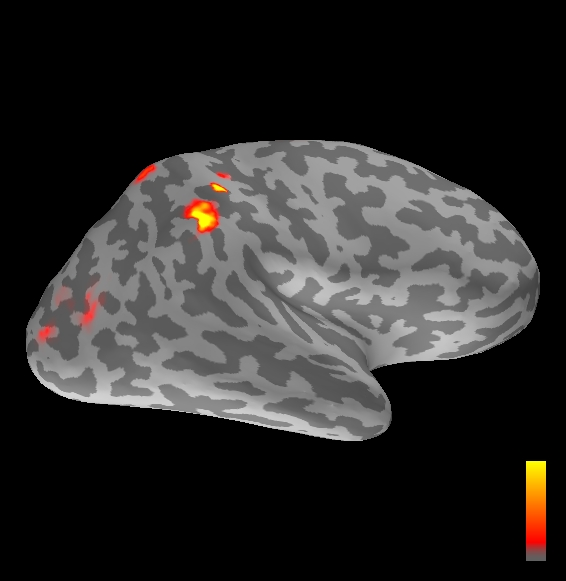}
\includegraphics[width=3cm]{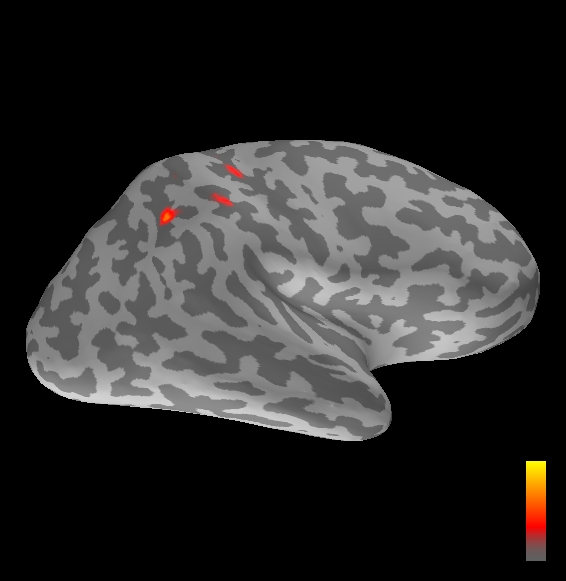}
\includegraphics[width=3cm]{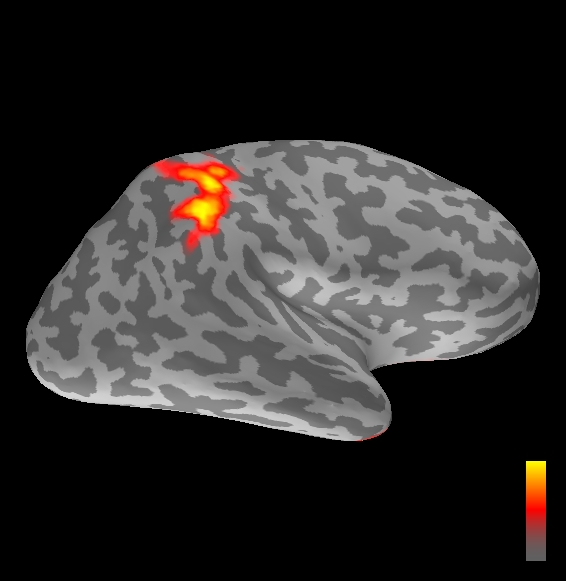}\\
\includegraphics[width=3cm]{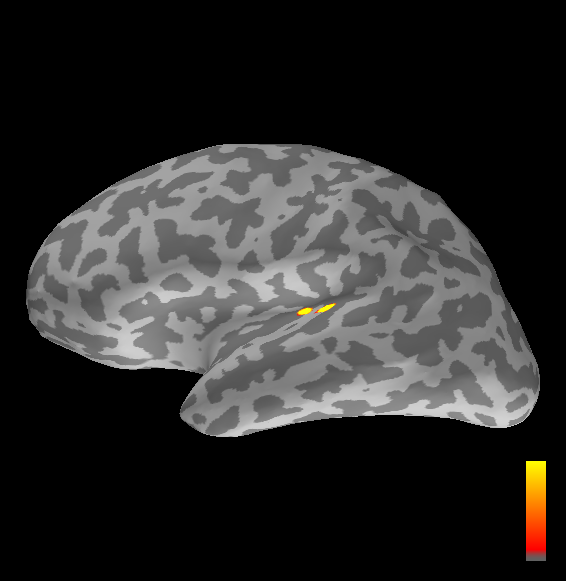}
\includegraphics[width=3cm]{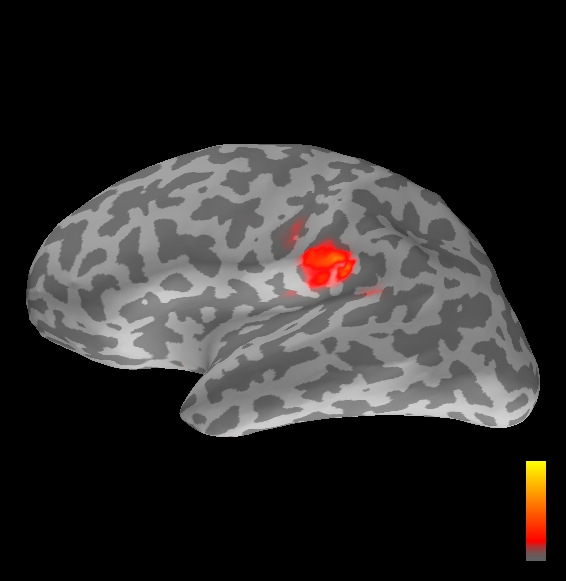}
\includegraphics[width=3cm]{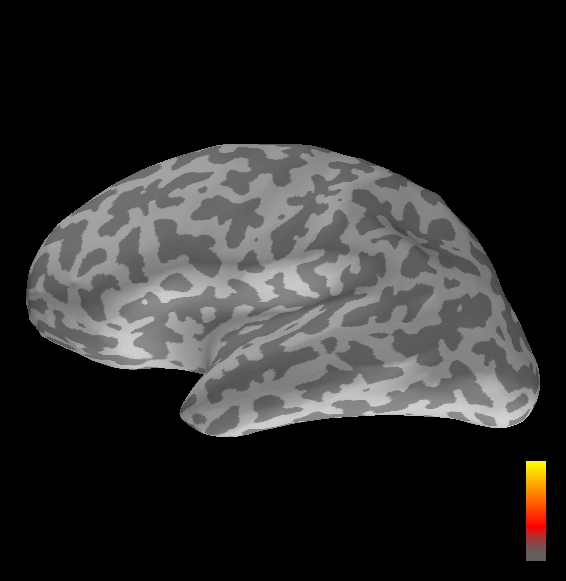}
\includegraphics[width=3cm]{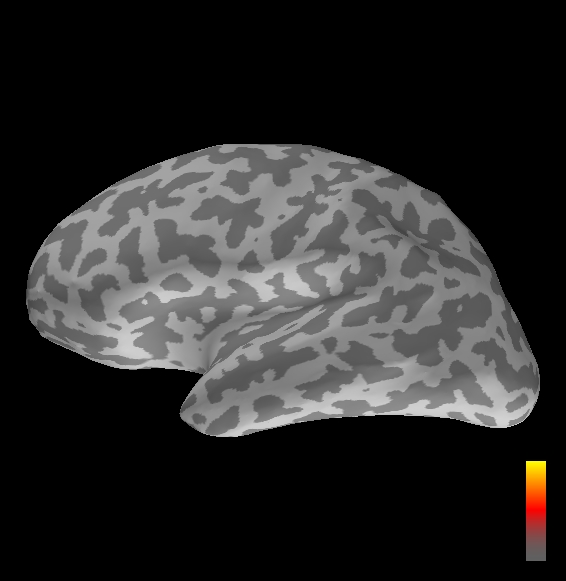}\\[4pt]
\includegraphics[width=3cm]{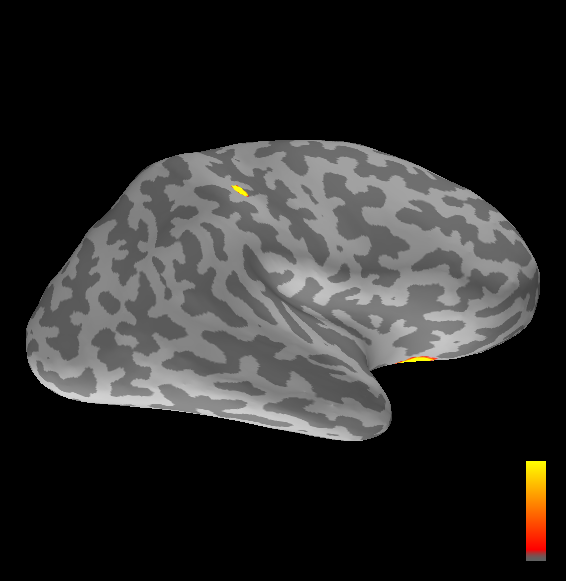}
\includegraphics[width=3cm]{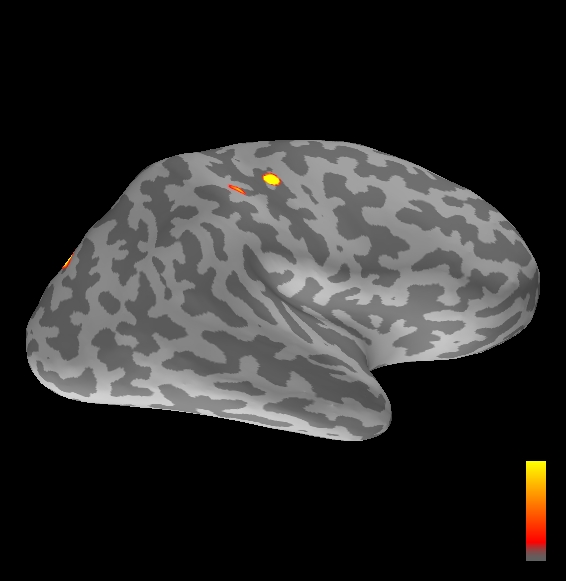}
\includegraphics[width=3cm]{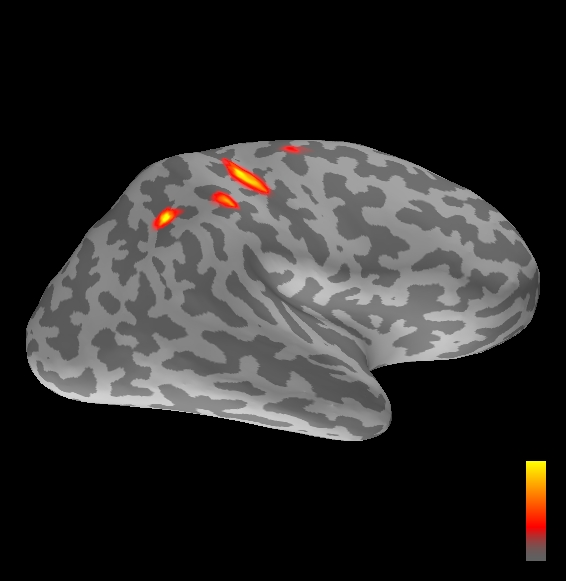}
\includegraphics[width=3cm]{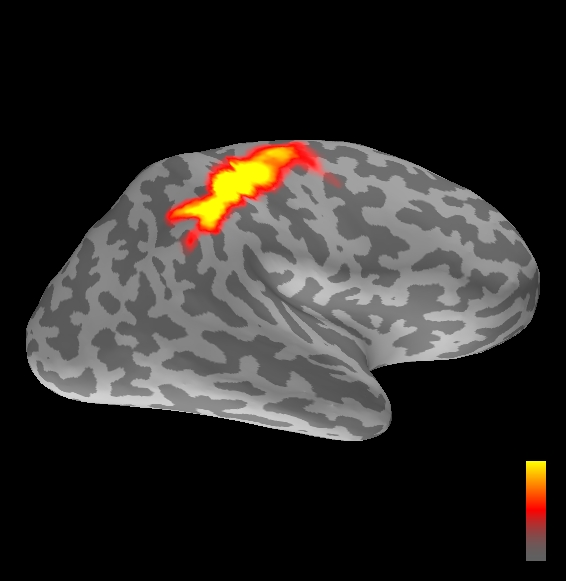}\\
\includegraphics[width=3cm]{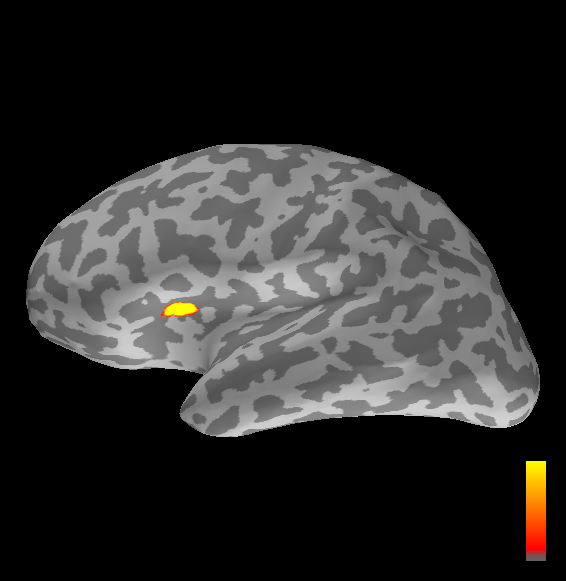}
\includegraphics[width=3cm]{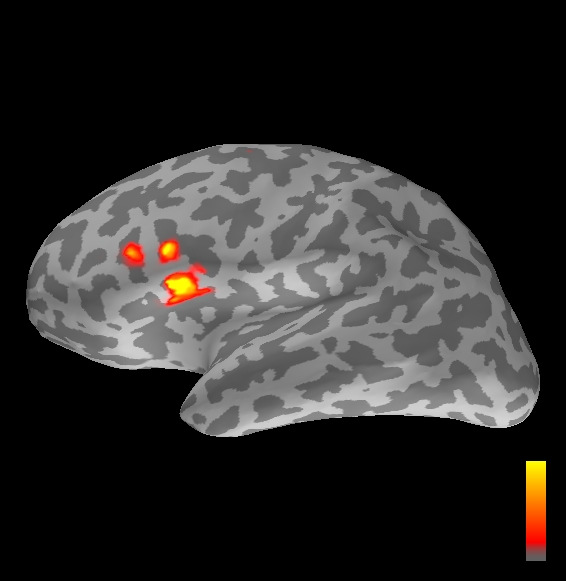}
\includegraphics[width=3cm]{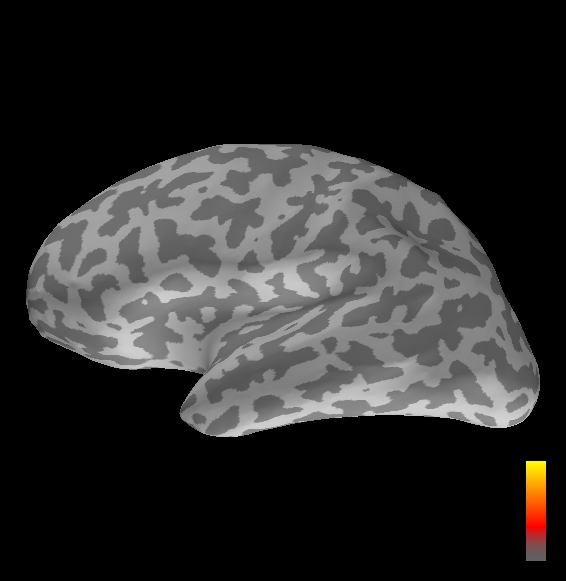}
\includegraphics[width=3cm]{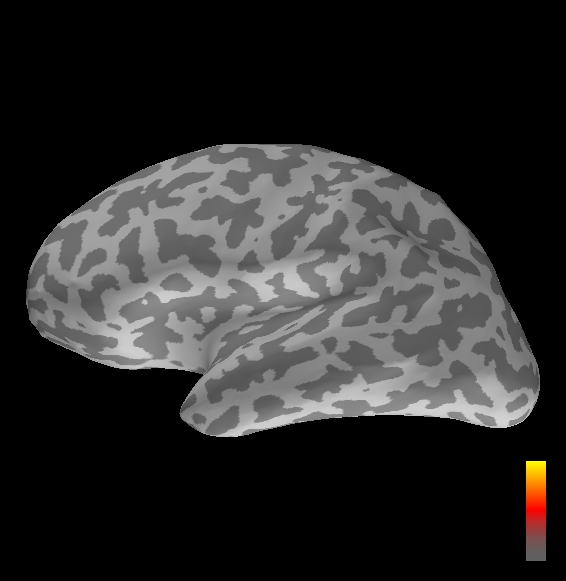}
\caption{Source estimates computed with the semi--analytic SMC (first column), the full SMC (second column), dSPM (third column) and SLORETA (fourth column).}
\label{fig:real_data_estimates}
\end{center}
\end{figure}

\begin{figure}
\begin{center}
\includegraphics[width=16cm]{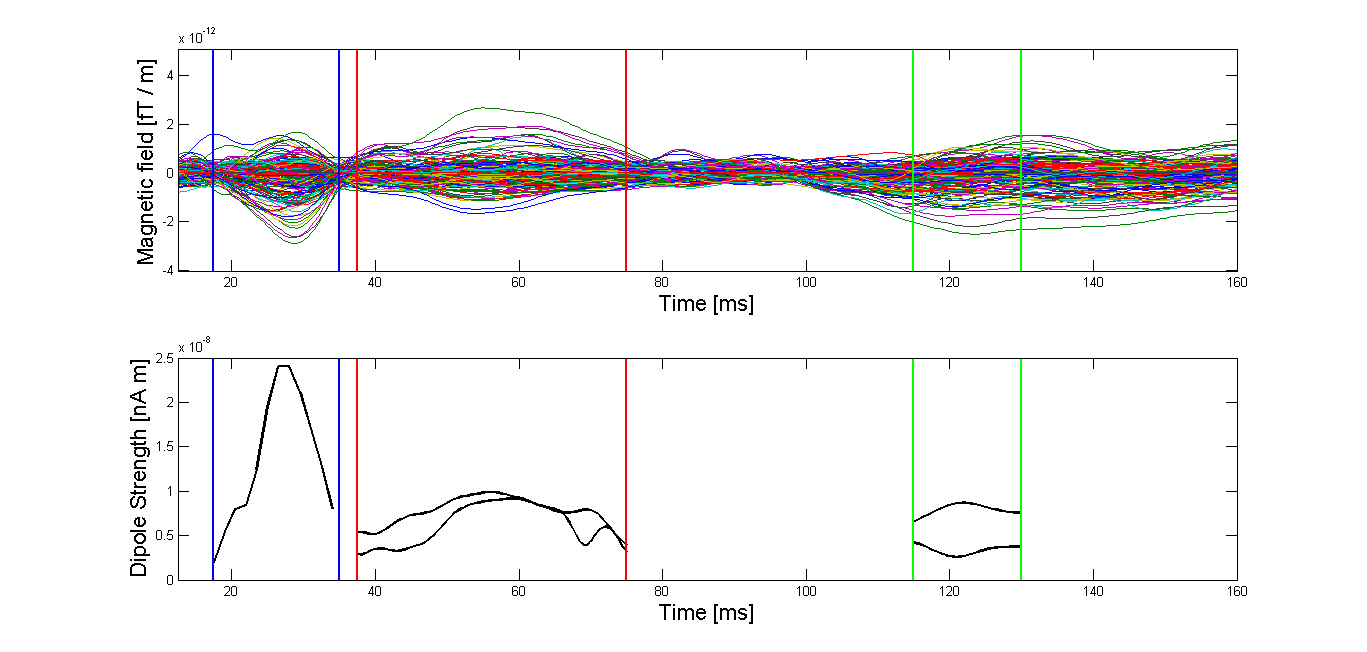}
\caption{Top panel: the MEG data, and the three temporal windows selected for the analysis. Bottom panel: estimated source time courses for the sources shown in Figure \ref{fig:real_data_estimates}.}
\label{fig:real_data_wf}
\end{center}
\end{figure}

In Figure \ref{fig:real_data_estimates} we show the estimated sources, superimposed on an inflated representation of the cortical surface:
light grey represents the gyri, dark grey the sulci, and color is used to represent either the posterior probability
(for the two SMCs) or the estimated activity (for dSPM and SLORETA). We remark that such visualization might be misleading, inasmuch the inflation
process tends to move apart closeby regions; hence, multiple blobs in the same area are often the consequence of nearby probable/active volumes.
We also recall that dSPM and SLORETA tend to provide widespread maps for relatively strong sources, and less diffused maps for relatively weak sources,
while the two SMC do the opposite, because when a source is strong the posterior probability is highly peaked and viceversa.
In light of this, the results of the semi--analytic SMC are in full agreement with those obtained by the full SMC, and seem to be richer than
those obtained by dSPM and SLORETA. Indeed, in the first row the localization of the primary somatosensory cortex is the same with all methods.
In the second time window/peak, the methods agree on the localization of the source in the right hemisphere;
the two SMC also estimate a source in the left hemisphere, which seems to be in accordance, for location and latency, with the literature on SEF responses \cite{maetal97}.
Similar considerations apply for the third time window/peak, where the primary somatosensory area is again active, and the two SMC localize an additional source
in the frontal region, confirmed by the literature.
In the two latter time windows, the semi--analytic SMC also finds, respectively, one and two additional sources in the central region of the brain; however, as
they are characterized by a lower probability and intensity, we are not showing them here.
The temporal waveforms of the sources estimated by the semi--analytic SMC are shown in Figure \ref{fig:real_data_wf}.

\section{Discussion}\label{par:discussion}

In this paper we have described how to use a recent class of Monte Carlo algorithms, called SMC samplers, for solving
Bayesian inverse problems described by a semi--linear structure, i.e. when the data depend linearly on a subset of the
unknowns and non--linearly on the remaining ones.

First we have shown that, under Gaussian assumptions for the likelihood and for the prior over the linear variables,
it is possible to compute a closed form of the marginal likelihood for the non--linear variables and for the conditional
posterior distribution for the linear ones. Therefore, the SMC sampler needs only to be applied to approximate the marginal posterior
for the non--linear variables, with the obvious advantage that part of the solution is computed analytically.
Second, we have observed that the sequence of distributions naturally chosen for the semi--analytic SMC does not coincide
with the sequence of the marginal distributions one would use in the full SMC. Instead, the distributions used in the first step of the semy--analytic SMC can be interpreted
as the marginals of the posterior densities of a Bayesian model in which both the noise standard deviation and the prior for
the linear and the non--linear variables change with the iterations. Therefore, the full SMC and the semi--analytic SMC
reach the same target distribution following different paths. From a theoretical perspective, it remains interesting
to extend the class of models to which we can apply this semi--analytic approach beyond just the class of Gaussian models.
For instance, one may introduce a hierarchical structure and choose a suitable
hyperprior for the parameters that are fixed a priori in our current approach, like the noise standard deviation and
the variance of the prior for the linear variables; for example, choosing a inverse--gamma hyperprior would lead to a
student-$t$ distribution. 

Then we have applied the semi--analytic method to the
inverse problem in MEG, where one wants to reconstruct the brain activity from the recorded magnetic field.
Here we have introduced several hypotheses about the model structure. First of all  we have used a multi--dipole model,
in which data depend linearly on the moments of the current dipoles, but not on their number and locations. In this context our
approach extends the work of \cite{soluar14}, where a full SMC was used to estimate multiple dipoles from a single spatial
distribution of the magnetic field: here, by exploiting the linearity with respect to the dipole moments, we can use
the semi--analytic SMC to estimate sources from a whole time--series, under the assumption that the number of dipoles
and their locations do not change in time. Such combined use of data at different time points is advantageous at least
for two reasons. On one hand, a time--series will contain more information about the source than a single time point,
and should therefore enable more accurate localization of the dipoles; in this respect, our approach is prefereable to the
use of several independent samplers, each one applied to a different time point. On the other hand, the computational
cost of the proposed algorithm is dominated by calculations that do not depend on the length of the time--window, thanks to the additional assumption
of independence between different time points. Alternative choices may include to model the linear variables as a Hidden Markov Model
and apply Kalman filtering/smoothing for their estimation, like in \cite{gimide13}.

We validated our method with different simulated experiments. First of all, we have shown that the semi--analytic
method produces
lower--variance approximations of the marginal distributions for the source locations, compared to those produced by a full SMC sampler.
This means that, the number of particles being equal, the approximations obtained by different runs of the semi--analytic SMC
are more similar to each other than those obtained by different runs of the full SMC. This was expected, as a natural consequence of the fact that we are sampling much fewer variables.

Then we have tested the behavior of the proposed algorithm with time--series generated by source configurations containing
between 2 and 4 dipoles, with different levels of temporal correlation. The results show that our approach is able to estimate with
high accuracy correlated as well as uncorrelated sources. Indeed, in all conditions the localization error remains of the order
of few millimeters. An indirect comparison with \cite{soluar14} suggests that we have improved in terms of localization accuracy,
as our localization error obtained from noisy data is lower than that obtained with noise--free data by the full SMC, particularly
for the more numerous configurations. We have also investigated the behaviour of the method with respect to a mis--specification of the prior.
We have shown that a mis--specification of the standard deviation for the dipole moment leads to a worsening of the performances. Namely
a too small value may lead to replacing a single true dipole with two estimated ones, while a too large value may lead to missing
a weak source. This suggests that the Gaussian distribution with fixed standard deviation might be a bit too restrictive:
possible future work may include adaptive estimation of this parameter, or the use of a different prior distribution, as suggested earlier
in this Section.  As a last numerical experiment, we have used simulations to show that the computational time of the algorithm does not depend on the length of
the time window given in input.

Finally, we have analyzed an experimental dataset recorded during somatosensory stimulation.
Here, we have observed that the estimates produced by the semi--analytic SMC are in full agreement with those obtained by the full SMC,
and provide in addition the time courses of the sources. The spatial maps produced by the semi--analytic SMC are slightly more focal, reasonably due to the fact
that a time window carries more information about the source location than a single time point. On the other hand, the new method also localizes
few additional sources in the middle of the brain, although with lower probability and intensity, that seem not to be due to physiological activity.
Possible explanations of this drawback include modeling errors, such as non perfectly stationary source locations, or the presence of spatial correlation
in the noise components.

Future research will be devoted to better asses the performance of the proposed method in real scenarios.
In addition, it would be helpful to decrease the computational cost by parallelizing the code with GPU programming.

\section*{Acknowledgements}

The authors have been supported in part by a grant of the INDAM--GNCS (Istituto Nazionale di Alta Matematica -- Gruppo Nazionale Calcolo Scientifico).

\section*{References}

\bibliography{biblio}

\begin{thebibliography}{10}

\bibitem{caetal08}
C.~Campi, A.~Pascarella, A.~Sorrentino, and M.~Piana.
\newblock A {R}ao-{B}lackwellized particle filter for magnetoencephalography.
\newblock {\em Inverse Problems}, 24:025023, 2008.

\bibitem{daetal00}
A.~Dale, A.K. Liu, B.R. Fischl, R.L. Buckner, J.W. Belliveau, J.D. Lewine, and
  E.~Halgren.
\newblock Dynamic statistical parametric mapping: Combining fmri and meg for
  high-resolution imaging of cortical activity.
\newblock {\em Neuron}, 26:55--67, 2000.

\bibitem{dafoka05}
G.~Dassios, A.S. Fokas, and F.~Kariotou.
\newblock On the non-uniqueness of the inverse {M}{E}{G} problem.
\newblock {\em Inverse Problems}, 21:L1--L5, 2005.

\bibitem{dedoja06}
P.~{Del~Moral}, A.~Doucet, and A.~Jasra.
\newblock Sequential {M}onte {C}arlo samplers.
\newblock {\em Journal of the Royal Statistical Society B}, 68:411--436, 2006.

\bibitem{dedoja12}
P.~{Del~Moral}, A.~Doucet, and A.~Jasra.
\newblock An adaptive sequential {M}onte {C}arlo method for approximate
  {B}ayesian computation.
\newblock {\em Statistics and Computing}, 22:1009--1020, 2012.

\bibitem{dogoan00}
A.~Doucet, S.~Godsill, and C.~Andrieu.
\newblock On sequential monte carlo sampling methods for bayesian filtering.
\newblock {\em Statistics and Computing}, 10:197--208, 2000.

\bibitem{dojo11}
A.~Doucet and A.M. Johansen.
\newblock A tutorial on particle filtering and smoothing: Fifteen years later.
\newblock In {\em The {O}xford {H}andbook of {N}onlinear {F}iltering}. Oxford
  University Press, 2011.

\bibitem{fokuma04}
A.S. Fokas, Y.~Kurylev, and V.~Marinakis.
\newblock The unique determination of neuronal currents in the brain via
  magnetoencephalography.
\newblock {\em Inverse Problems}, 20:1067--1082, 2004.

\bibitem{gimide13}
F.~Giraud, P.~Minvielle, and P.~Del Moral.
\newblock Advanced interacting sequential {M}onte {C}arlo sampling for inverse
  scattering.
\newblock {\em Inverse Problems}, 29:095014, 2013.

\bibitem{gr95}
P.J. Green.
\newblock Reversible jump {M}arkov {C}hain {M}onte {C}arlo computation and
  {B}ayesian model determination.
\newblock {\em Biometrika}, 82:711--732, 1995.

\bibitem{haetal93}
M.~{H\"{a}m\"{a}l\"{a}inen}, R.~Hari, J.~Knuutila, and O.V. Lounasmaa.
\newblock Magnetoencephalography: theory, instrumentation and applications to
  non-invasive studies of the working human brain.
\newblock {\em Reviews of Modern Physics}, 65:413--498, 1993.

\bibitem{juetal05}
S.C. Jun, J.S. George, J.~Par\'e-Blagoev, S.M. Plis, D.M. Ranken, D.M. Schmidt,
  and C.C. Wood.
\newblock Spatiotemporal bayesian inference dipole analysis for {M}{E}{G}
  neuroimaging data.
\newblock {\em NeuroImage}, 28:84--98, 2005.

\bibitem{juetal06}
S.C. Jun, J.S. George, S.M. Plis, D.M. Ranken, D.M. Schmidt, and C.C. Wood.
\newblock Improving source detection and separation in a spatiotemporal
  bayesian inference dipole analysis.
\newblock {\em Physics in Medicine and Biology}, 51:2395--2414, 2006.

\bibitem{kigeve83}
S.~Kirkpatrick, C.D. Gelatt, and M.P. Vecchi.
\newblock Optimization by simulated annealing.
\newblock {\em Science}, 220:671--680, 1983.

\bibitem{maetal97}
F.~Mauguiere, I.~Merlet, N.~Forss, S.~Vanni, V.~Jousmaki, P.~Adeleine, and
  R.~Hari.
\newblock Activation of a distributed somatosensory cortical network in the
  human brain. {A} dipole modelling study of magnetic fields evoked by median
  nerve stimulation. part {I}: location and activation timing of {S}{E}{F}
  sources.
\newblock {\em Electroencephalography and Clinical Neurophysiology},
  104:281--289, 1997.

\bibitem{mole99}
J.C. Mosher and R.M. Leahy.
\newblock Source localization using {R}ecursively {A}pplied and {P}rojected
  ({R}{A}{P}) {M}{U}{S}{I}{C}.
\newblock {\em IEEE Transactions on Signal Processing}, 47:332--340, 1999.

\bibitem{molele92}
J.C. Mosher, P.S. Lewis, and R.M. Leahy.
\newblock Multiple dipole modeling and localization from spatio-temporal
  {M}{E}{G} data.
\newblock {\em IEEE Transactions on Biomedical Engineering}, 39:541--557, 1992.

\bibitem{naasjo12}
C.F.H. Nam, J.A.D. Aston, and A.M. Johansen.
\newblock Quantifying the uncertainty in change points.
\newblock {\em Journal of Time Series Analysis}, 33:807--823, 2012.

\bibitem{pa02}
R.M. Pascual-Marqui.
\newblock Standardize low resolution electromagnetic tomography
  (s{L}{O}{R}{E}{T}{A}: technical details.
\newblock {\em Methods and Findings in Experimental and Clinical Pharmacology},
  24:5--12, 2002.

\bibitem{savela07}
S.~Sarkka, A.~Vehtari, and J.~Lampinen.
\newblock Rao-blackwellized particle filter for multiple target tracking.
\newblock {\em Information Fusion}, 8:2--15, 2007.

\bibitem{sa87}
J.~Sarvas.
\newblock Basic mathematical and electromagnetic concepts of the biomagnetic
  inverse problem.
\newblock {\em Phys. Med. Biol.}, 32:11--22, 1987.

\bibitem{scvovo08}
D.~Schuhmacher, B.T. Vo, and B.N. Vo.
\newblock A consistent metric for performance evaluation of multi-object
  filters.
\newblock {\em IEEE Transactions on Signal Processing}, 56:3447--3457, 2008.

\bibitem{seetal02}
K.~Sekihara, S.S. Nagarajan, D.~Poeppel, A.~Marantz, and Y.~Miyashita.
\newblock Application of an meg eigenspace beamformer to reconstructing
  spatio-temporal activities of neural sources.
\newblock {\em Human Brain Mapping}, 15:199--215, 2002.

\bibitem{soka04}
E.~Somersalo and J.P. Kaipio.
\newblock {\em Statistical and computational inverse problems}.
\newblock Springer Verlag, 2004.

\bibitem{soetal13}
A.~Sorrentino, A.M. Johansen, J.A.D. Aston, T.E. Nichols, and W.S. Kendall.
\newblock Dynamic filtering of static dipoles in {M}agnetoencephalography.
\newblock {\em Annals of Applied Statistics}, 7:955--988, 2013.

\bibitem{soluar14}
A.~Sorrentino, G.~Luria, and R.~Aramini.
\newblock Bayesian multi-dipole modeling of a single topography in meg by
  adaptive sequential monte-carlo samplers.
\newblock {\em Inverse Problems}, 30:045010, 2014.

\bibitem{tava82}
A.~Tarantola and B.~Valette.
\newblock Inverse problems = quest for information.
\newblock {\em Journal of Geophysics}, 50:159--170, 1982.

\bibitem{vvetal97}
B.D. {Van~Veen}, W.~van Drongelen, M.~Yuchtman, and A.~Suzuki.
\newblock Localization of brain electrical activity via linearly constrained
  minimum variance spatial filtering.
\newblock {\em IEEE Transactions on Biomedical Engineering}, 44:867--880, 1997.

\bibitem{vi07}
M.~Vihola.
\newblock Rao-blackwellised particle filtering in random set multitarget
  tracking.
\newblock {\em IEEE Transactions on Aerospace and Electronic Systems},
  43:689--705, 2007.

\bibitem{waza11}
J.~Wan and N.~Zabaras.
\newblock A {B}ayesian approach to multiscale inverse problems using the
  sequential {M}onte {C}arlo method.
\newblock {\em Inverse Problems}, 27, 2011.

\bibitem{zhjoas13arxiv}
Y.~Zhou, A.M.Johansen, and J.A.D. Aston.
\newblock Towards automatic model comparison, an adaptive sequential monte
  carlo approach.
\newblock {\em arXiv}, stat.ME:1303:3123, 2013.

\end{thebibliography}

\end{document}